\documentclass[aps,prx,showpacs,amsmath,amssymb,amsfonts,lengthcheck,twocolumn,footinbib,floatfix,superscriptaddress]{revtex4-1}
\usepackage[utf8]{inputenc}
\usepackage{amsmath,amssymb}
\usepackage[colorlinks=true,citecolor=blue,linkcolor=blue,urlcolor=blue]{hyperref}%
\usepackage{url}
\usepackage{graphicx}
\usepackage{scrextend}
\usepackage{xcolor}
\usepackage{cleveref}
\usepackage{soul}
\usepackage[normalem]{ulem} 

\newcommand{\ma}[1]{#1_{\mathcal{A}}}
\newcommand{\mb}[1]{#1_{\mathcal{B}}}

\definecolor{orange}{rgb}{1,0.5,0}
\newcommand{\jf}[1]{{\bf\color{orange}{#1}}}

\newcommand{\rev}[1]{#1}

\begin{document}

\title{Phase separation and nucleation in mixtures of particles with different temperatures}
\author{Efe Ilker}
\affiliation{Physico-Chimie Curie UMR 168, Institut Curie, PSL Research University, 26 rue d'Ulm, 75248 Paris Cedex 05, France}
\author{Jean-Fran\c{c}ois Joanny}
\affiliation{Physico-Chimie Curie UMR 168, Institut Curie, PSL Research University, 26 rue d'Ulm, 75248 Paris Cedex 05, France}
\affiliation{Coll\'{e}ge de France, 11 place Marcelin Berthelot, 75005 Paris, France}
\begin{abstract}
\rev{Differences in activities in colloidal particles are sufficient to drive phase separation between active and passive (or less active) particles, even if they have only excluded volume interactions.} In this \rev{paper, we study the phase 
separation kinetics and propose a theory of phase separation of colloidal mixtures in the diffusive limit.} \rev{Our model considers a mixture of diffusing particles coupled to different thermostats, it thus has a non-equilibrium nature due to the temperature differences.} 
However, we show that indeed the system recovers \rev{ an effective equilibrium thermodynamics in the dilute limit.} We  obtain phase diagrams showing \rev{the} asymmetry in concentrations due to activity differences. By using a more general approach, we 
show the equivalence of phase separation kinetics with the well known Cahn-Hilliard theory. On the other hand, higher order expansions \rev{in concentration} indicate the emergence of non-equilibrium effects leading to \rev{a breakdown of the equilibrium analogy.}  We lay out the general theory in terms of accessible parameters which we 
demonstrate by several applications. In this simple formalism, we capture \rev{a positive surface tension for hard spheres}, and interesting scaling laws for interfacial properties, droplet growth dynamics, and 
phase segregation conditions. \rev{Several of our results are in agreement with existing numerical simulations while we also propose testable predictions.} 
\end{abstract}
\maketitle

\section{Introduction}
Many-body systems self-organize and exhibit macroscopic collective behavior, 
which is amenable to coarse-grained dynamics \cite{chaikin1995}. The principles of 
equilibrium thermodynamics which provide the essential tools for studying 
equilibrium self-assembly  are not adapted to  the new phenomena emerging in 
active systems \cite{marchetti2013}. While active systems are non-equilibrium 
in their nature, with each constituent having its own energy budget and 
objective, it is quite fascinating how these systems can display distinct 
characteristics which are sometimes non-trivially related to the known 
basic physical 
principles. A typical example is that of active phase separation.

In living organisms, activity-driven phase separation plays an important role by 
promoting self-organization and increasing the efficiency of biological functions 
inside cells \cite{zwicker2014,woodruff2017}, which are due to operate in a 
very crowded and noisy environment. The constant use of energy in unequal 
amounts by individual cellular sub-components reflects their varieties in dynamical and 
chemical activities \cite{agudo2019} \rev{as well as} with their innate physiological 
differences. In turn, these activity differences may enhance phase separation 
by creating effective attractions between alike components. Remarkably, similar 
characteristics and governing principles are observed at various length scales 
on diverse biological systems and they are now also guiding synthetic model systems
\cite{palacci2013, singh2017,zhang2018,lin2018,popescu2018}. 

It is now well known that activity differences in colloidal particles are sufficient to drive phase separation between active and passive (or less active) hard spheres having only excluded volume interactions. These systems can be 
considered to have particles with two different temperatures which mimics the effect of activity differences \cite{grosberg2015,weber2016,tanaka2017}. In principle, the constituent particles exchange energy, the hotter ones 
providing extra energy for the colder ones.  This, in turn, indicates a non-equilibrium behavior \cite{grosberg2018}. Similarly, in polymeric systems, tiny activity differences in active/passive polymer mixtures enhance phase
segregation \cite{smrek2017,smrek2018}. If the system is diffusive, the effective temperature can simply be deduced from the effective diffusivity. For instance, at room temperature $T$, bacteria may reach an effective 
temperature $\ma{T}\sim 10^2-10^3 T$ in translational motion by chemotaxis \cite{berg2008coli}. These realizations are not limited to biological or polymeric systems, but \rev{are} also seen in plasma physics \cite{pitaevskii2012} and in the description of 
thermal phases in the interstellar medium \cite{wolfire2003, cox2005}. Due to having different heating/cooling processes, ionized hydrogen reaches a temperature $\sim 10^2$ times larger than atomic hydrogen distributed in the interstellar medium \cite{ferriere2001interstellar}. In many other contexts, the concept of effective temperature \cite{cugliandolo2011} \rev{proves quite} useful in understanding large-scale 
phenomena that originate from microscopic motility or even chemical differences between the building components \cite{exartier1999,crisanti2012,chertovich2004,pande2000}. Considering the growing interest in these systems, it is important to study and establish a theory of phase separation starting from the microscopic dynamics.

In this work,  we extend the theory of mixtures of particles with different temperatures introduced in 
Ref.\cite{grosberg2015} to inhomogeneous mixtures and we study the phase separation kinetics of 
solutions composed of two diffusing species, which are coupled to different 
thermostats. Starting from the microscopic model (Sec.\ref{sec2}), we derive and present the theory in the dilute limit (Secs.\ref{sec2}-\ref{sec4}) and obtain the scaling laws for phase behaviors. 
Even though the system has inherently non-equilibrium properties due to 
activity differences, it obeys an 
effective equilibrium thermodynamics in the dilute limit including the interfacial contributions, which generalizes the Cahn-Hilliard theory of solutions \cite{cahn1958} to two temperature mixtures. \rev{We verify this comparing the equilibrium thermodynamic route with the mechanical one.} In Section \ref{sec5}, we demonstrate example applications of the theory in diverse systems, \rev{and we discuss the results in comparison to previous studies with new predictions.} In Section \ref{sec6}, we consider \rev{higher} order corrections in concentrations which break the equilibrium analogy and raise the 
difficulty to define a scalar temperature. 
 For purely hard-core interactions, we present higher order corrections 
in density 
to the theory by considering depletion interactions between spheres of different 
activities, and obtain results, which do not seem to be easily obtainable otherwise. \rev{We summarize and discuss all these results in the conclusion.}

\section{Microscopic model}\label{sec2}
We start by introducing the microscopic model \rev{and derive} an effective 
thermodynamic description, which is valid when the solutions that we study are 
very dilute. We follow the lines of Ref.\cite{grosberg2015} \rev{but we extend this approach to inhomogeneous mixtures. Thus, we} study a solution 
of mixed particles satisfying overdamped Langevin equations, in contact with 
reservoirs at different temperatures
\begin{eqnarray}
\zeta_{m} \dot{x}_{m}=-\partial_{m} U + (2 T_{m} \zeta_{m})^{1/2} \xi_{m} (t)
\end{eqnarray}
where  $U$ is the overall interaction potential between the particles. The 
position of particle $m$ is denoted by $x_{m}$, its friction coefficient by $
\zeta_{m}$, its temperature by $T_{m}$ and $\xi_{m}(t)$ is a standard zero mean, 
unit variance, Gaussian white noise.
Here, we consider two different species of particles each being in contact with a 
thermostat at temperature $\ma{T}$ or $\mb{T}$. This dynamics presented as a  
Langevin equation for each particle in the system, can be reformulated as a 
Fokker-Planck equation for a multi-particle probability distribution $P(\{{\bf r}\})$ 
where $\{{\bf r}\}$ is a vector whose components are the positions of all the 
particles. The Fokker-Planck equation is written in terms of the fluxes $J_{m}$ 
for each particle:
\begin{eqnarray}\label{FPmulti}
\frac{\partial P(\{{\bf r}\})}{\partial t}&=& -\sum_{m} \partial_{m} J_{m}, 
\nonumber\\
J_{m}&=&-\partial_{m} U P/\zeta_{m} - T_{m}\partial_{m} P /\zeta_{m}.
\end{eqnarray}
Distinguishing the two species of particles $\mathcal{A}$ and $\mathcal{B}$ 
and integrating over all coordinates except for one, we obtain the single particle 
distributions $p_{\alpha}({\bf r}_1)$ where now $\alpha,\beta=\mathcal{A}$ or $
\mathcal{B}$  \cite{grosberg2015}:
\begin{eqnarray}\label{FP1}
\frac{\partial p_{\alpha}{({\bf r}_1)}}{\partial t} =& \frac{T_{\alpha}}{\zeta_{\alpha}} \nabla _{{\bf r}_1}  ^2 p _{\alpha}({\bf r}_1)\rev{-
\frac{1}{\zeta_{\alpha}} 
\partial _{{\bf r}_1} p _{\alpha}({\bf r}_1)
\sum_{\beta} \bar{f}_{\alpha \beta}} ,\nonumber
 \\\nonumber\\
&\bar{f}_{\alpha\beta} = - N_{\beta}\int \frac{\partial u_{\alpha \beta}}{\partial {\bf r}_1} 
\frac{p_2 ^{\alpha\beta} ({\bf r}_1,{\bf r}_2)}{p _{\alpha}({\bf r}_1)} 
d {\bf r}_2 .
 \end{eqnarray}
and $N_{\beta}$ is the number of $\beta$-type particles. We may write the two 
particle densities in terms of single particle distributions and a pair distribution 
function  $p_2 ^{\alpha \beta} ({\bf r}_1, {\bf r}_2) = p_{\alpha} ({\bf r}_1)  
p_{\beta} ({\bf r}_2) g_{\alpha\beta} ({\bf r}_1 - {\bf r}_2)$. Accordingly, $ 
g_{\alpha\beta} ({\bf r}_1 - {\bf r}_2)$ is the pair distribution function and in the 
long time limit, it reaches a steady-state value, i.e., $ g_{\alpha\beta} ({\bf r}_1 - 
{\bf r}_2)\rightarrow  g_{\alpha\beta} ^{ss} ({\bf r}_1 - {\bf r}_2)$. The knowledge 
of the steady state pair distribution function is sufficient to close this set of 
equations. In general, $g_{\alpha\beta} ^{ss} ({\bf r}_1 - {\bf r}_2)$ depends on 
the particle concentrations and can be expanded in powers of the 
concentrations. In the dilute limit, when considering only pair interactions, the 
solution of Eq.~\eqref{FPmulti} imposes:
\begin{eqnarray}\label{eqgss}
g_{\alpha\beta} ^{ss} ({\bf r}_1 - {\bf r}_2) = \exp [ -u_{\alpha\beta} ({\bf r}_1 - 
{\bf r}_2)/T_{\alpha\beta}]
\end{eqnarray}
where  $T_{\alpha\beta}$ are the mobility-weighted average temperatures and 
are defined as $T_{\mathcal{AA}}\equiv T_{\mathcal{A}}$, $T_{\mathcal{BB}}
\equiv T_{\mathcal{B}}$, and $T_{\mathcal{AB}}=\left(\zeta_{\mathcal{B}}
T_{\mathcal{A}}+\zeta_{\mathcal{A}}T_{\mathcal{B}}\right)/(\zeta_{\mathcal{A}}
+\zeta_{\mathcal{B}} )$. 

In order to determine the forces $\bar{f}_{\alpha\beta}$, we rewrite the two-
particle distribution function as 
$p_2 ^{\alpha\beta} ({\bf r}_1, {\bf r}) = p_ {\alpha} ({\bf r}_1)  p_{\beta} ({\bf r}
_1+{\bf r})
\exp [ -u^{\alpha\beta} ({\bf r})/T_{\alpha\beta}]$ by defining ${\bf r}_2={\bf r}_1+
{\bf r}$. This change of variables 
inside the integrals yields
\begin{equation}
\begin{split}
\bar{f}_{\alpha\beta}= &\rev{-N_{\beta} T_{\alpha\beta}}\\& \times \int  \frac{\partial }{\partial 
{\bf r}_1}  \left(1-e^{-u_{\alpha\beta}({\bf r})/T_{\alpha\beta}}\right)   p_{\beta} 
({\bf r}_1+{\bf r}) d {\bf r} .
\end{split}
\end{equation}
Assuming that the concentrations vary slowly over \rev{a length scale
much larger} than the range of the pairwise interactions, we expand $p_{\alpha} 
({\bf r}_1+{\bf r})\approx p_{\alpha} ({\bf r}_1)+ {\bf r} \cdot \nabla  p_{\alpha} 
({\bf r}_1)+ \frac{1}{2}({\bf r} \cdot \nabla)^2  p_{\alpha} ({\bf r}_1)$.
While inserting in the integrand, the first term gives the uniform or 
homogeneous contribution to the force, the second term vanishes, and the third 
term provides an inhomogeneous contribution to the force. 

\subsection{Effective thermodynamic identities}
\label{efftherm}
We introduce the concentrations $c_{\alpha} ({\bf x})=N_{\alpha} p_{\alpha} ({\bf x})$, and obtain closed equations for the concentrations:
\begin{eqnarray}\label{cform1}
\frac{\partial c_{\alpha} ({\bf r}_1)}{\partial t}= \frac{T_{\alpha}}{\zeta_{\alpha}}\nabla_{{\bf r}_1} ^2 c_{\alpha}({\bf r}_1)-\frac{1}{\zeta_{\alpha}}\nabla_{{\bf r}_1} c_{\alpha}({\bf r}_1)\bar{f}_{\alpha}({\bf r}_1)
\end{eqnarray}
We have defined here the total mean force $\bar{f}_{\alpha}= \sum_{\beta}\bar{f}_{\alpha\beta}$ acting on  a particle of species $\alpha$ due to all the other particles. In this particular case, this total mean force is the 
gradient of a potential and we can write the conservation equation for the concentrations in the  Cahn-Hilliard form:
\begin{equation}\label{cform2}
\frac{\partial c_{\alpha} ({\bf r}_1)}{\partial t}=\frac{1}{\zeta_{\alpha}} \nabla_{{\bf r}_1}\cdot c_{\alpha}({\bf r}_1)\nabla_{{\bf r}_1} \mu_{\alpha},
\end{equation}
This equation defines the functions $\mu_{\alpha}$ as non-equilibrium analogs of chemical potentials. 
\begin{equation}
\mu_{\alpha}=\mu_{\alpha}^{\text{id}}+\Phi_{\alpha}, \hspace{10pt} \mu_{\alpha}^{\text{id}}=T_{\alpha}\ln c_{\alpha},\hspace{10pt} -\bar{f}_{\alpha}({\bf r}_1) = \nabla_{{\bf r}_1} \Phi_{\alpha}.
\label{chempotinteract}
\end{equation}
We decompose the non-equilibrium chemical potentials as sums of a homogeneous part, which depends only on the concentration and a non-homogeneous part, which depends on the concentration gradients. 
\begin{subequations}\label{mueq}
\begin{eqnarray}
\mu_{\alpha} &=&\mu_{\alpha} ^{0}+\mu_{\alpha} ^{\nabla}, \\
\mu_{\alpha} ^{0}&=&T_{\alpha} \ln c_{\alpha} + \sum_{\beta}T_{\alpha\beta} B_{\alpha\beta} c_{\beta}, \\
\mu_{\alpha} ^{\nabla}&=&\sum_{\beta}T_{\alpha\beta} \Lambda_{\alpha\beta} \nabla ^2 c_{\beta}, 
\end{eqnarray}
\end{subequations}
The quantities $B_{\alpha\beta}=\int  (1-e^{-u^{\alpha\beta}({\bf r})/T_{\alpha\beta}}) d {\bf r}$ are the effective excluded volumes or \rev{second virial coefficients \footnote{\rev{This differs from the standard definition of virial coefficients by a factor of 2. We chose it this way to keep the free energy in Flory-Huggins form.} }} and $\Lambda_{\alpha\beta}=\frac{1}{6}\int  r^2 
(1-e^{-u^{\alpha\beta}({\bf r})/T_{\alpha\beta}}) d {\bf r}$.

The non-equilibrium chemical potentials can themselves be calculated as the functional derivatives of an effective non-equilibrium free energy $\mu_{\alpha}=\delta \mathcal{F}/\delta c_{\alpha}$ which is the functional 
derivative of the total free energy $\mathcal{F}[\ma{c},\mb{c}]=\int f d\bf{r}$ with respect to the concentration $c_{\alpha}({\bf r})$. The reconstruction of free energy from the chemical potentials gives the free energy 
per unit volume which is given by
\begin{subequations}
\begin{eqnarray}
\label{freeenergy}
f &=&f^{0}+f ^{\nabla},\\
f^{0}&=&\sum_{\alpha}T_{\alpha} c_{\alpha} \ln (c_{\alpha}/e)+\sum_{\alpha,\beta}\frac{1}{2}T_{\alpha\beta}B_{\alpha\beta}c_{\alpha} c_{\beta}\label{fb}, \\
f^{\nabla}&=&\sum_{\alpha,\beta}\frac{1}{2}L_{\alpha\beta}(\nabla c_{\alpha})(\nabla c_{\beta}) 
\end{eqnarray}
\end{subequations}
where $L_{\alpha\beta}=-T_{\alpha\beta}\Lambda_{\alpha\beta}$  is negative. 
The free energy $f^0$ for uniform concentrations has already been derived in 
Ref.\cite{grosberg2015}. It has a Flory-
Huggins form with differences in interactions dictated by the two different 
temperatures and the effective excluded 
volumes $B_{\alpha\beta}$'s (second virial coefficients). The contrast in 
temperatures further enhances the 
tendency toward demixing that is inherent to the Flory-Huggins free energy. For 
instance, in the case where all the \rev{virial coefficients}
$B_{\alpha}$ are identical ($\ma{B}=B_{\mathcal{AB}}=\mb{B}$), (which 
corresponds in particular to equal-sized 
hard spheres), solely the difference in temperatures  drives asymmetrically 
weighted interactions between 
particles that are responsible for phase separation. A crucial 
remark is that the friction 
coefficients $\zeta_{\alpha}$ or the mobilities, which are the inverse of the 
friction coefficients, only enter through 
the effective pairwise temperature $T_{\mathcal{AB}}$ which becomes indistinct 
for $\ma{T}=\mb{T}$. Hence, a difference 
in diffusivities $\mathcal{D}_{\alpha}\propto T_{\alpha}/\zeta_{\alpha}$ at the 
same temperature $\ma{T}=\mb{T}$ 
but $\ma{\zeta}\neq\mb{\zeta}$ has no influence on the thermodynamics. This 
is expected a priori since in this 
case the system is at thermal equilibrium. 

It is remarkable that the concept of effective free energy can be extended to 
inhomogeneous solutions of particles with two different temperatures. In \rev{a} dilute limit, at lowest order in the 
concentration gradients, this shows compatibility with the Landau-Ginzburg theory \cite{bray2002}. A major difference in this case is that the steady-state is maintained at the expense of extra power input \cite{oono1998}. 

\subsection{Internal stress}\label{sec2b}
In order to derive the local stress tensor $\sigma_{ij}$ in the solution, we first calculate it from the 
free energy as could be done in an equilibrium system and then give a 
mechanical derivation based on the Irving-Kirkwood formulation of the stress, 
which leads to the same results.  

\paragraph{Effective thermodynamic construction:}
In a deformation  of the 
volume $V+\delta V$, the total work is obtained by integrating the work 
across each surface element. If we call $dS_j$ the surface element and $u_i$ the infinitesimal displacement along respectively $i-$ and $j$-direction in 
Cartesian coordinates, the work associated to the deformation of the volume is
\begin{eqnarray}
\delta W= \int _{\partial V} dS_j u_i \sigma_{ij}
\end{eqnarray}
On the other hand, if there is an effective free energy, the work done by \rev{the} 
displacement is $\delta W= \delta \mathcal{F}$ . Hence, we may obtain the 
stress tensor $\sigma_{ij}$ from  the surface contribution to the change in free 
energy  $\delta \mathcal{F}$  which can be written as:
\begin{eqnarray}
\delta \mathcal{F}= \int _{\delta V} f d{\bf r} +\int \delta f d{\bf r}.
\end{eqnarray}
The first term on the right-hand side gives the surface integral $\int f ({\bf u 
\cdot dS})$ while the second one can be converted to a surface integral by 
expanding $\delta f$ in terms of of the changes in the concentrations  $\delta c_{\alpha}$ induced 
by the deformation, i.e.,
\begin{equation}
\begin{split}
\delta f=&\mu_{\mathcal{A}} \delta c_{\mathcal{A}} +\mu_{\mathcal{B}} \delta 
c_{\mathcal{B}}\\&+\nabla \cdot \left(\frac{\partial f}{\partial (\nabla 
c_{\mathcal{A}})} \delta c_{\mathcal{A}} \right)+\nabla \cdot \left(\frac{\partial f}
{\partial (\nabla c_{\mathcal{B}})} \delta c_{\mathcal{B}} \right)
\end{split}
\end{equation}
We study here the stress in a steady state, and as discussed in the previous 
sections, the chemical potentials 
$\mu_{\mathcal{A}}$ and $\mu_{\mathcal{B}}$  are constant throughout the 
volume. We can therefore eliminate the changes in concentrations $\delta 
c_{\alpha}$ by using the conservation of the total numbers of particles $
\mathcal A$ and $\mathcal B$ during the deformation $\int_V \delta c_{\alpha}  
+ \int_{\delta V} c_{\alpha}  =0 $. 

The total change in the free energy can then be written as a surface integral
\begin{equation}
\begin{split}
\delta \mathcal{F}= \int _{\partial V} dS_j  &\bigg[u_i\left(f-\mu_{\mathcal{A}}
c_{\mathcal{A}}-\mu_{\mathcal{B}}c_{\mathcal{B}}\right) \delta_{ij} \\
&+\delta c_{\mathcal{A}} \left(L_{\mathcal{A}} \partial_i c_{\mathcal{A}} + L_{\mathcal{AB}} \partial_i c_{\mathcal{B}} \right)\\&+ \delta c_{\mathcal{B}} \left(L_{\mathcal{B}} \partial_i c_{\mathcal{B}} + L_{\mathcal{AB}} \partial_i c_{\mathcal{A}} \right) \bigg]
\end{split}
\end{equation}
Finally, the variation of the concentration on the surface is given by  $\delta c_{\alpha}= -{\bf u} \cdot  \nabla c_{\alpha}$ for both species.
The resulting stress tensor reads:
\begin{equation}\label{stresseq}
\begin{split}
\sigma_{ij}=&\left(f-\mu_{\mathcal{A}}c_{\mathcal{A}}-\mu_{\mathcal{B}}
c_{\mathcal{B}}\right) \delta_{ij}\\& - \partial_i c_{\mathcal{A}}\left[ L_{\mathcal{A}} 
\partial_j  c_{\mathcal{A}} + L_{\mathcal{AB}} \partial_j  c_{\mathcal{B}} \right]\\&- 
\partial_i c_{\mathcal{B}}\left[ L_{\mathcal{B}} \partial_j  c_{\mathcal{B}}+ 
L_{\mathcal{AB}} \partial_j  c_{\mathcal{A}} \right].
\end{split}
\end{equation}
Accordingly, the pressure $p$ can be deduced directly  from the diagonal 
component of the stress, $p \delta_{ii}=-\sigma_{ii}$ which gives in three 
dimensions:
\begin{equation}
\begin{split}
p=&\left(\mu_{\mathcal{A}}c_{\mathcal{A}}+\mu_{\mathcal{B}}c_{\mathcal{B}}-f
\right) \\&+
\frac{1}{3} \left( L_{\mathcal{A}} (\nabla c_{\mathcal{A}} )^2 + L_{\mathcal{B}} 
(\nabla c_{\mathcal{B}} )^2  +2L_{\mathcal{AB}} (\nabla c_{\mathcal{A}} )
(\nabla c_{\mathcal{B}}) \right)
\end{split}
\end{equation}
The first term contains the locally uniform  pressure $p^0=
\left(\mu^0_{\mathcal{A}}c_{\mathcal{A}}+\mu^0_{\mathcal{B}}c_{\mathcal{B}}-
f^0
\right)$ that is given by the standard Gibbs-Duhem equation and the gradient 
terms including the contributions of $\mu^{\nabla}_{\mathcal{A}}$, $\mu^{\nabla}_{\mathcal{B}}$ and $f^{\nabla}$ determine the interfacial contributions.

\paragraph{Irving-Kirkwood method:}An alternative, more general method to calculate the stress tensor without any reference to the equilibrium thermodynamics, has been proposed by Irving and Kirkwood \cite{irving1950}, 
starting from the mechanical virial equation \cite{hansen1995}. The stress tensor $\sigma_{ij}^{(v)}$ is given by:
\begin{eqnarray}
\label{irving}
\sigma_{ij} ^{(v)}= \sigma_{ij} ^{K}+\underbrace{ \frac{1}{2}\bigg \langle \sum_{\alpha,\beta}\frac{r_i r_j}{{\bf r}} \left(\frac{\partial u_{\alpha\beta} ({\bf r})}{\partial r} \right) \bigg \rangle  }_{\sigma_{ij}^{u}}
\end{eqnarray}
where $\sigma_{ij} ^{K}$ is the stress in the absence of interactions (for  an 
ideal gas), while the second part is the contribution to the stress due to 
interparticle potentials that we name $\sigma_{ij}^{u}$. The interaction part of 
the stress  can be decomposed into a sum over the particle species $\alpha$ 
and $\beta$ and the average in Eq.\eqref{irving} 
can be calculated using the two-particle probability distribution $p_2^{\alpha\beta}({\bf r_1},{\bf r})$.  As in the previous 
paragraphs, we expand the two-particle densities around ${\bf r_1}$, and  
make a change of variables to calculate the average values. This leads to the 
following stress,
\begin{equation}\label{ikstress}
\begin{split}
\sigma_{ij} ^{u} =   &\sum_{\alpha,\beta}\frac{N_{\alpha} 
N_{\beta}}{2}\\& \times\int \frac{r_i r_j}{{\bf r}} \frac{\partial u_{\alpha\beta} ({\bf r})}{\partial 
r} \sum_{l=1} ^{\infty} \frac{(-{\bf r}\cdot \nabla)^{l-1} }{ l !} p_2 ^{\alpha\beta} 
({\bf r}_1,{\bf r}) d {\bf r}
\end{split}
\end{equation}
The rest of the calculation is straightforward. We  give the full result of this 
calculation in Appendix \ref{appb}. It turns out that the stress tensor calculated by the 
Irving-Kirkwood method is different \rev{from} the stress calculated from the free 
energy; it can be written as $\sigma_{ij} ^{(v)}=\sigma_{ij} ^{(f)}+ \sigma_{ij} '$ 
where $\sigma_{ij} ^{(f)}$ is the stress obtained in the previous paragraph from 
the free energy Eq.~\eqref{stresseq}. This shows that the stress is not 
defined in a unique way \cite{schofield1982}. However, it conserves all the 
properties of $\sigma_{ij} ^{(f)}$ for our analysis, and strictly does not alter the 
force-balance since $\partial_i \sigma_{ij} ' =0$.  Interestingly, 
$\sigma_{ij} ^{(v)}$, could just be obtained from a free energy 
perturbation by adding surface 
terms $-\frac{1}{3}\nabla^2 \left[  L_{\mathcal{A}} c_{\mathcal{A}}^2 + 2 L_{\mathcal{AB}}  c_{\mathcal{A}} 
c_{\mathcal{B}} + L_{\mathcal{B}} c_{\mathcal{B}}^2 \right]$ 
to the 
free energy given by Eq.~\eqref{freeenergy}. 

This result validates that in the limit of low 
concentrations, we can still use the thermodynamic approach to 
calculate the stress inside the solution. In the more general case where 
there is no effective free energy, one would need to rely on the Irving-Kirkwood description.
A final note is that an alternative formulation of Irving-Kirkwood method can be 
achieved by using microscopic force-balance \cite{aerov2014,kruger2018} in 
the time evolution equations Eqns~\eqref{cform1}, thus summing up all mean 
internal forces.

\section{Phase lines and the critical point}\label{sec3}
We now use the effective thermodynamic description of the solution to calculate the phase diagram of a solution of particles at two different temperatures. 
\subsection{\rev{Dimensionless effective thermodynamic quantities}}Let us introduce first the volume fractions $\phi_{\alpha}=c_{\alpha} B_{\alpha}/\epsilon_{\alpha}$ where $\epsilon_{\alpha}\equiv 
B_{\alpha}/v_{\alpha}$ is the conversion factor to molecular volume $v_{\alpha}$. The volume factions are 
well defined only if the total volume fraction is smaller than one. They must then satisfy $\ma{\phi}+\mb{\phi}\leq 1$. We also define $\mb{\beta}=\mb{\epsilon}B_{\mathcal{AB}}/(\mb{B})$, the temperature ratio 
$\alpha_T=\ma{T}/\mb{T}$, volume ratio $\alpha_v=\ma{v}/\mb{v}$,  and the friction ratio $\alpha_{\zeta}=\ma{\zeta}/\mb{\zeta}$. Finally, we define $\hat{L}_{\alpha\beta}= 
\mb{T}^{-1}L_{\alpha\beta}\ma{v}/(v_{\alpha}v_{\beta})$. Accordingly, we set the dimensionless free energy density $\hat{f}=\mb{T}^{-1}\ma{v} f$ while the total free energy \rev{is} $\hat{\mathcal{F}}=\int \hat{f} d{\bf r}$. As a 
result, we obtain the \rev{dimensionless} chemical potentials $\hat{\mu}_{\alpha}=\delta  \hat{\mathcal{F}}/ \delta \phi_{\alpha}$:

\begin{align}
\ma{\hat{\mu}}\equiv \mb{T}^{-1}\mu_{\mathcal{A}}=
        \begin{aligned}[t] &\alpha_T\left(\ln \phi_{\mathcal{A}} + \ma{\epsilon}\phi_{\mathcal{A}}\right)+\frac{\alpha_{T}+\alpha_{\zeta}}{1+\alpha_{\zeta}}\mb{\beta} \phi_{\mathcal{B}}\\&-\ma{\hat{L}}\nabla^2 
        \ma{\phi}-\hat{L}_{\mathcal{AB}}\nabla^2 \mb{\phi},
 \end{aligned}\\
\mb{\hat{\mu}}\equiv \mb{T}^{-1}\alpha_v\mu_{\mathcal{B}}= \begin{aligned}[t] &\alpha_v\left(\ln \phi_{\mathcal{B}} + \mb{\epsilon}\phi_{\mathcal{B}}\right)+\frac{\alpha_{T}+\alpha_{\zeta}}{1+\alpha_{\zeta}} 
\mb{\beta}\phi_{\mathcal{A}}\\&-\mb{\hat{L}}\nabla^2 \mb{\phi}-\hat{L}_{\mathcal{AB}}\nabla^2 \ma{\phi}, \end{aligned}
\end{align}
where we ignored the density-independent terms. Note that $\mb{\hat{\mu}}$ has an extra scaling factor $\alpha_v$ in order to conserve all the functional properties to construct the thermodynamic functions of the previous 
section. Similarly for  pressure we have $\hat{p}=\mb{T}^{-1}\ma{v} p$. This completes our transformation to dimensionless functionals $X[\ma{c}({\bf r}),\mb{c}({\bf r})]\rightarrow 
\hat{X}[\ma{\phi}({\bf r}),\mb{\phi}({\bf r})] $. As in the previous case, we can separate these into locally uniform and interfacial \rev{components}, i.e., $\hat{X}=\hat{X}^0+\hat{X}^{\nabla}$.

\subsection{Two phase coexistence} At zero-flux steady state for single particle concentrations, we should have uniform chemical potentials and pressure. For a mixed state this would suggest to have uniform concentrations (a single phase). However, if there are any two phases coexisting, they should satisfy the following conditions at their interface:
\begin{subequations}\label{eqco}
\begin{eqnarray}
\hat{\mu}_{\mathcal{A}}^0\left( \phi_{\mathcal{A}}^{a}, \phi_{\mathcal{B}}^{a} \right)&=&\hat{\mu}_{\mathcal{A}}^0\left(\phi_{\mathcal{A}}^{b}, \phi_{\mathcal{B}}^{b} \right),\\
\hat{\mu}_{\mathcal{B}}^0\left( \phi_{\mathcal{A}}^{a}, \phi_{\mathcal{B}}^{a} \right)&=&\hat{\mu}_{\mathcal{B}}^0\left(\phi_{\mathcal{A}}^{b}, \phi_{\mathcal{B}}^{b} \right),\\
\hat{p}^0\left( \phi_{\mathcal{A}}^{a}, \phi_{\mathcal{B}}^{a} \right)&=&\hat{p}^0\left(\phi_{\mathcal{A}}^{b}, \phi_{\mathcal{B}}^{b} \right),
\end{eqnarray}
\end{subequations}
where $a$ and $b$ denote the two coexisting phases. Together, these suggest no net particle exchange and force-balance at phase boundary while concentrations continuously vary from one phase to the other. We obtain the 
coexistence curve (or binodal line) in our phase diagrams by numerically solving the above conditions.

\subsection{Spinodal line}
The stability of the uniform state $\ma{\phi}({\bf r})=\ma{\phi}^0$ and  $\mb{\phi}({\bf r})=\mb{\phi}^0$ can be analyzed by linearizing $\partial \ma{\phi} / \partial t$ and $\partial \mb{\phi} / \partial t$ around the 
uniform state by introducing $\ma{\phi}({\bf r})=\ma{\phi}^0+\delta \ma{\phi}({\bf r})$ and $\mb{\phi}({\bf r})=\mb{\phi}^0+\delta \mb{\phi}({\bf r})$ where $\ma{\phi}^0$ and $\mb{\phi}^0$ denote the uniform states. As a 
result, we obtain in Fourier space the equation of the relaxation of a perturbation of wave vector ${\bf q}$, $ {\bf{\delta \tilde{\phi}}}({\bf q})=(\delta \tilde{\phi}_{\mathcal{A}}({\bf q}),
\delta \tilde{\phi}_{\mathcal{B}}({\bf q}))$
\begin{equation}\label{gameq1}
\frac{\partial {\bf{\delta \tilde{\phi}}}({\bf q})}{\partial t}=-q^2\Gamma{\bf{\delta \tilde{\phi}}}({\bf q}),
 \end{equation}
 with the relaxation matrix:
 \begin{eqnarray}\label{gameq2}
 \Gamma&=&
\mb{T}\left(
 \begin{matrix}
 \ma{\phi}^0\ma{\zeta}^{-1} &   0 \\
 0 &   \alpha_v^{-1}\mb{\phi}^0\mb{\zeta}^{-1}
 \end{matrix}
  \right) \kappa^{-1}_p,  \nonumber\\ \nonumber\\
  \kappa^{-1}_p &=& \left(\begin{matrix}
 \alpha_T \left(\frac{1+\ma{\epsilon}\ma{\phi}}{\ma{\phi}} \right) &   \frac{\alpha_{T}+\alpha_{\zeta}}{1+\alpha_{\zeta}}\mb{\beta} \\
 \frac{\alpha_{T}+\alpha_{\zeta}}{1+\alpha_{\zeta}}\mb{\beta} &   \alpha_v \left(\frac{1+\mb{\epsilon}\mb{\phi}}{\mb{\phi}} \right) 
 \end{matrix}\right)
 \label{fluceq}
 \end{eqnarray}
where $\kappa^{-1}_p$ is the inverse of the compressibility matrix obtained from the linearization of the chemical potentials and we have neglected terms of order $q^4$. Note that in $\kappa^{-1}_p$ and for the remainder, we 
no longer write the superscript zero of $\phi_{\alpha}$'s. The instability occurs when at least one of the eigenvalues of $\Gamma$ becomes negative. Thus, it is enough to determine the limit where the determinant $ 
|\kappa^{-1}_p|$ is negative. The non-dimensionalized spinodal line equation is obtained as 
\begin{eqnarray}
\label{spindim}
\nu_s=\frac{1+\ma{\epsilon}
\phi_{\mathcal{A}}}{\phi_{\mathcal{A}}} \frac{1+\mb{\epsilon}\phi_{\mathcal{B}}}{\phi_{\mathcal{B}}}-\frac{(\alpha_{T}+
\alpha_{\zeta})^2\mb{\beta}^2}{(1+\alpha_{\zeta})^2\alpha_T \alpha_v }
=0
\end{eqnarray}
for vanishing \rev{wavevector} q. The instability occurs when $\nu_s<0$. The spinodal line \rev{is} 
symmetric if $\ma{\epsilon}=\mb{\epsilon}$. It is clear 
that a larger contrast of activity, i.e., a larger $\alpha_T$ \rev{enlarges} unstable region of the phase diagram.

\subsection{Critical point}\label{sec3d}
The critical point is calculated by finding the point where the two phases in 
equilibrium are identical \cite{gibbs1878}. This is the point along spinodal line where the 
fluctuations are maximum. Hence, we search for the 
point along the spinodal line where the gradient of the spinodal line in the 
volume fraction parameter space $\nabla \nu_s={\bf \widehat{ \phi}}
_{\mathcal{A}}\partial_{\ma{\phi}} \nu_s + {\bf \widehat{  \phi}}_{\mathcal{B}}
\partial_{\mb{\phi}} \nu_s$ is aligned with  the eigenvector ${\bf e}_0$ of the 
inverse compressibility matrix, corresponding to the eigenvalue $\epsilon=0$. 
Accordingly, the volume fractions at the critical point $(\ma{\phi}^*,\mb{\phi}^*)$ 
satisfy:
\begin{subequations}\label{fxdpt}
\begin{eqnarray}\frac{\phi_{\mathcal{B}}^* (1+\mb{\epsilon}\phi_{\mathcal{B}}^*)}{(1+\ma{\epsilon}\phi_{\mathcal{A}}^*)^2}=\frac{\alpha_T(1+\alpha_{\zeta})}{\mb{\beta}(\alpha_{T}+\alpha_{\zeta})},\\ \nonumber \\
\nu_s\left(\ma{\phi}^*,\mb{\phi}^*\right)=0.
\end{eqnarray}
\end{subequations}
We observe that even at $\ma{\epsilon}=\mb{\epsilon}$ where the 
spinodal line is symmetric, the location of the critical point can be shifted 
along the spinodal line by controlling the ratio $\alpha_T/ \alpha_v$. One easy way to see that is to use the conjugate of Eq.\eqref{fxdpt} and symmetrize these two forms for $\ma{\phi}$ and $\mb{\phi}$. 
Accordingly, choosing $\alpha_T>\alpha_v$ moves the critical point toward the $
\mathcal{B}$-rich part of the phase diagram while setting $\alpha_T<
\alpha_v$  moves it toward the  $\mathcal{A}$-rich side. However, if $\ma{\epsilon}\neq\mb{\epsilon}$, it seems more complicated to get a sense on such symmetrization and one should follow Eq.\eqref{thetac}. We will 
investigate further the properties of such asymmetric phase diagrams in 
Section \ref{sec5} in order to get a hint on the structure of coexisting phases that 
can be either both liquid-like phases or a solid-like and a gas-like phase.

\subsection{Condition for existence of phase separation}\label{sec3e}
In the previous paragraphs, we outlined the calculation of  phase diagrams using 
the effective free energy obtained in the limit of low concentrations, but 
imposing no constraints on the volume fractions and 
assuming that both phases remain fluid. If the volume fractions are high 
enough in one of the phases, this phase 
cannot be fluid and is solid. There is in this case equilibrium between 
a liquid or gaseous (very dilute) phase and a solid phase for which our 
concentration 
expansion is not accurate but approximate. Still, it qualitatively indicates a crystalline phase. Simulations show that this phase exhibits hexagonal packing in 2-dimensions \cite{weber2016}, while both face-centered cubic and hexagonal-close packed structures in 3-dimensions \cite{chari2019}.  On the other hand, for hard-sphere interactions and 
considering that the two phases remain fluid, we improve our approximation in 
Section \ref{sec6} by adding one order in concentration. 

A general constraint on the volume fractions is that the total volume 
fraction is 
not space filling and that it is smaller than the critical concentration 
for space filling $\phi_{max}$ (for a random packing of identical spheres $\phi_{max}\simeq 0.64$). Here, for generality, we stick with $\phi_{max}=1$ following the general literature. This choice, together with low density 
approximation is sufficient to observe qualitative tendency to phase-separate while varying interaction and activity parameters.

A common 
approach to study the conditions for phase separation \cite{lebowitz1964} is to impose this 
space filling condition for the spinodal line given by Eq.~\eqref{spindim}. This conjecture accepts 
the emergence of an instability region in the phase diagram as the 
sufficient condition for phase separation. An alternative more restrictive view would be that the critical point 
should exist inside a 
physical phase diagram. In this case, the valid condition is the existence of the 
critical point inside the physical regime. These two approaches are 
identical if the critical point is at the tip of the spinodal line while the latter 
condition delays the onset of the coexistence region. We evaluate 
both conditions for the cases we consider (Fig. \ref{fig3}).

\section{Phase separation kinetics}\label{sec4}
Knowing the coexisting phases and the effective thermodynamic 
description, we can develop the theory of phase separation kinetics for 
two 
temperature mixtures. We start here by determining the surface tension 
between two phases at equilibrium and then discuss phase separation 
kinetics. Note that the interface between the two phases is 
stable only if the surface tension is positive.

\begin{figure}
  \centering \includegraphics[width=0.85\columnwidth]{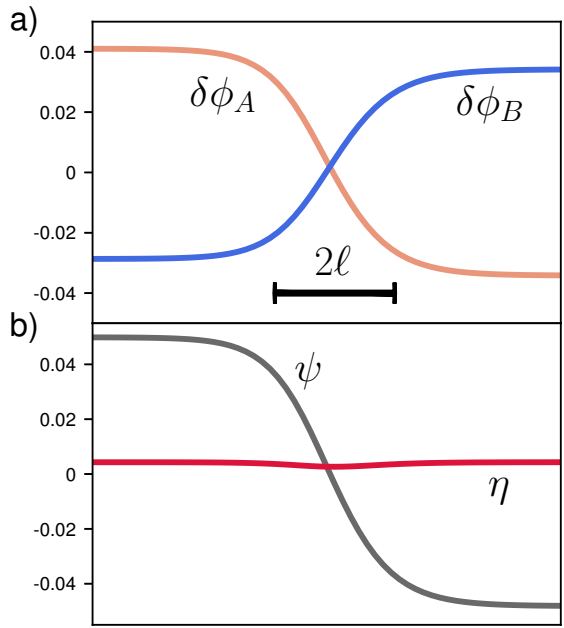}
  \caption{Example density profiles between coexisting phases for hard-spheres with $\alpha_v=27$ and $\alpha_T=20$ and $\ma{\epsilon}=\mb{\epsilon}=\mb{\beta}=8$ and hence $\tan\theta^*\approx 0.836$. We show a) profiles \rev{of the
  particle volume fractions } $\delta\phi_{\mathcal{A}}=\phi_{\mathcal{A}}(z)-\phi_{\mathcal{A}}^*$ and  $\delta\phi_{\mathcal{B}}=\phi_{\mathcal{B}}(z)-\phi_{\mathcal{B}}^*$, b) profiles \rev{of the normal  
  coordinates $\psi(z)$ and $\eta(z)$;  we observe that} $\eta(z)\ll \psi(z)$. The interface width is $2\ell$ and away from the interface the densities take constant values at two coexisting phases. }\label{fig1}
\end{figure}
\subsection{Surface tension}
We consider a mixture with two phases at equilibrium and with a flat interface 
between the two phases. In this geometry, the two concentrations or the 
volume fractions vary only along one direction, say the $z$-direction so that  $
\partial_y c_{\alpha}= \partial_x c_{\alpha}=0$. The stress is isotropic in the bulk 
of each phase but becomes anisotropic close to the interface.  The interfacial 
tension between the two phases can be calculated from the stress distribution 
in the solution
\begin{eqnarray}
\label{gameq}
\gamma=\int _{z_{a}} ^{z_{b}} (\sigma_{xx}-\sigma_{zz}) dz
\end{eqnarray}
where the integration is from one phase (phase-$a$) to the other (phase-$b$). Using our calculated value of the stress tensor $\sigma_{ij}$, Eq.~\eqref{stresseq}, we see that the isotropic component of the stress proportional
to $\delta_{ij}$ cancels out and that only the non-diagonal components of the stress contribute to the surface tension: they vanish  for $\sigma_{xx}$ but do not vanish for $\sigma_{zz}$.  In dimensionless form,  
$\hat{\sigma}_{ij}\equiv\ma{v} \mb{T}^{-1}\sigma_{ij}$, we find

\begin{align}
\hat{\sigma}_{xx}-\hat{\sigma}_{zz}= \hat{L}_{\mathcal{A}} (\partial_z \phi_{\mathcal{A}})^2+\hat{L}_{\mathcal{B}} (\partial_z \phi_{\mathcal{B}})^2\nonumber\\+2\hat{L}_{\mathcal{AB}} (\partial_z \phi_{\mathcal{A}})( 
\partial_z \phi_{\mathcal{B}})
\label{surftens}
\end{align}
In order to determine the surface tension $\gamma$ from Eqns~\eqref{gameq} and ~\eqref{surftens}, we then need to determine the concentration profiles along $z$-direction. We first introduce the boundary conditions in the two 
phases in equilibrium: i) $\hat{\mu}_{\mathcal{A}}\left( \phi^{a}_{\mathcal{A}}, \phi^{a}_{\mathcal{B}} \right)=\hat{\mu}_{\mathcal{A}}^{\dagger}=\hat{\mu}_{\mathcal{A}}\left( \phi^{b}_{\mathcal{A}}, \phi_{\mathcal{B}}^{b} 
\right)$, ii) $\hat{\mu}_{\mathcal{B}}\left( \phi_{\mathcal{A}}^{a}, \phi_{\mathcal{B}}^{a} \right)=\hat{\mu}_{\mathcal{B}}^{\dagger}=\hat{\mu}_{\mathcal{B}}\left( \phi_{\mathcal{A}}^{b}, \phi^{b}_{\mathcal{B}} \right)$ and 
iii) $\hat{p}\left( \phi^{a}_{\mathcal{A}}, \phi^{a}_{\mathcal{B}} \right)=\hat{p}^{\dagger}=\hat{p}\left(\phi_{\mathcal{A}}^{b}, \phi_{\mathcal{B}}^{b} \right)$ where the values with daggers are the constant values of the 
chemical potentials and the pressure at equilibrium. The concentration profiles can then be calculated from the coupled differential equations:
\begin{subequations}
\begin{eqnarray}
\hat{\mu}_{\mathcal{A}} ^0(z)-\hat{\mu}_{\mathcal{A}} ^{\dagger}=\hat{L}_{\mathcal{A}} \nabla^2 \phi_{\mathcal{A}}(z) + \hat{L}_{\mathcal{AB}}\nabla^2 \phi_{\mathcal{B}}(z),\\
\hat{\mu}_{\mathcal{B}} ^0(z)-\hat{\mu}_{\mathcal{B}} ^{\dagger}=\hat{L}_{\mathcal{B}} \nabla^2 \phi_{\mathcal{B}}(z) + \hat{L}_{\mathcal{AB}}\nabla^2 \phi_{\mathcal{A}}(z).
\end{eqnarray}\label{profile}
\end{subequations}
Using the Gibbs-Duhem expression of the free energy in the phases at equilibrium, $\hat{f}^{0}=\hat{\mu}_{\mathcal{A}}^0\phi_{\mathcal{A}}+\hat{\mu}_{\mathcal{B}}^0\phi_{\mathcal{B}}-\hat{p}^0$ and 
integrating Eq.\eqref{profile} is consistent with $\hat{\sigma}_{xx}-\hat{\sigma}_{zz}=\Delta \hat{f}\left[\phi_{\mathcal{A}}(z),\phi_{\mathcal{B}}(z)\right]$ where the tilted free energy is given by:
\begin{eqnarray}
\Delta \hat{f}= \hat{f}-\left(\hat{\mu}_{\mathcal{A}}^{\dagger}\phi_{\mathcal{A}}+\hat{\mu}_{\mathcal{B}}^{\dagger}\phi_{\mathcal{B}}-\hat{p}^{\dagger}\right).\label{delforiginal}
\end{eqnarray}
While this is the generic form, the same treatment more specifically implies that $\Delta \hat f=2\Delta \hat {f}^0$.
The tilted free energy is the difference between the local free energy along the 
concentration profiles and the energy obtained from the so-called common 
tangent construction. Since the common tangent construction gives the 
minimal possible free energy,  the tilted free energy is always positive  $\Delta 
\hat{f} >0$ \rev{as long as one can solve  the set of equations \eqref{profile} with appropriate boundary conditions}. We may therefore write an alternative form of the interfacial tension 
:
\begin{eqnarray}
\hat{\gamma}\equiv\mb{T}^{-1}\ma{v}\gamma =2\int_{z_{a}}^{z_{b}} \Delta \hat{f}^0\left(\phi_{\mathcal{A}}(z),\phi_{\mathcal{B}}(z)\right) dz.\label{gameqfull}
\end{eqnarray}
\rev{If there is a consistent profile, t}he surface tension  $\gamma$ is therefore always positive and the 
interface between the two phases is stable. The set of equations \eqref{profile}
can be solved numerically by linearization of the two equations around one 
boundary (say 
phase $a$)  and integrating up to the other boundary (phase $b$) using a 
shooting method. Alternatively, 
an analytical approximation can be obtained by considering the system close to 
the critical point as done in the next paragraph.
This analytical approximation is in excellent agreement with the numerical results.

\textit{Surface tension near the critical point:}
In the vicinity of the critical point, we show in Appendix \ref{normal} how the effective thermodynamics can be expressed as a function of a single order  parameter $\psi$ which is a linear combination volume fractions 
relative 
to the critical point. In addition, the other normal coordinate $\eta$ gives the normal distance from the critical point, and a phase separation occurs when a solution exists $\eta_a\approx \eta_b>0$.  Each value of $\eta_a$ 
defines the two coexisting phases $\psi_a$ and $\psi_b$.The transformed coordinates are illustrated in Fig.\ref{fig1}, where at first the concentration profiles $\ma{\phi}(z)$ and $\mb{\phi}(z)$ are obtained by solving Eq.
\eqref{profile} numerically.
The effective tilted free energy density is given in \eqref{totalf} as a function of the order parameter only, which is obtained by transforming \eqref{delforiginal} to normal coordinates. Minimization with respect to $\psi$ \rev{leads to }
\begin{equation}
 \Delta \hat  f^0=\frac 1 4 {k_{\psi}} (\psi-\psi_{a})^2 (\psi-\psi_b)^2=\frac 1 2 \hat L_{\psi} ({\bf \nabla} \psi)^2.
 \label{totalf1}
 \end{equation}
\rev{Eqs.\eqref{gameqfull},\eqref{totalf1} suggest that $\hat{\gamma}= \hat{L}_{\psi}\int 
(\nabla\psi)^2$, and hence the sign of $L_{\psi}$ determines the sign of the surface tension. It is then simple to prove that the surface tension for equal-sized hard spheres is positive for all $\alpha_T$ values. For 
hard spheres with varying size ratios, we} \rev{checked by numerical evaluation of $L_{\psi}$ that
it is positive for the values of the parameters $\alpha_V$ and $\alpha_T$ that lead to a phase
separation except when $\alpha_v \ll 1$ or $\alpha_v \gg \alpha_T$ (see Appendix \ref{appd} for calculations, and Section \ref{abp} for more discussion)}.
The solution of \eqref{totalf1} gives the concentration profile
 \begin{equation}
  \psi = \frac{\Delta \psi_{ab}}2 \tanh z/\ell
 \end{equation}
 where the interface width is  $\ell= \left(\frac{8L_{\psi}}{{\Delta \psi_{ab}} ^2 
 k_{\psi}}\right)^{1/2}$.
The surface tension can then be calculated by integration of Eq.\eqref{surftens} where we keep 
only the terms involving the gradient of $\psi$
\begin{equation}\label{gammasc1}
 \hat \gamma \approx \frac{1}{12} (\Delta \psi_{ab}) ^3 \left( 2 k_{\psi} \hat{L}_{\psi} \right)^{1/2}.
 \end{equation}

 In order to look at the scaling variation with parameters of the mixture such as $\alpha_T$ or 
 $\alpha_v$, one must express the dimensionless quantities that we used as functions of these 
 parameters. Taking as an example equal-sized hard spheres where $\alpha_v=1$, and hence 
 $\alpha_{\zeta}=1$, $\ma{\epsilon}=\mb{\epsilon}=\mb{\beta}=8$, the volume fractions at the 
 critical point are given by $\ma{\phi}^*=\alpha_T ^{-1}$, $\mb{\phi}^*=1/8+(5/4)\alpha_T^{-1}$ 
 when $\alpha_T \gg 1$. Considering a particle mixture of given volume fractions of particles 
 (which could be \rev{called} the laboratory conditions) $\ma{\phi}^0$ and $\mb{\phi}^0$ \rev{defined 
 by $\phi_{\alpha}^0=V^{-1}\int_V \phi_{\alpha}({\bf r}) d{\bf r}$ and for finite volume fractions in the vicinity of the critical point, we obtain:}
\begin{equation}
  \hat \gamma \sim  (\ma{\phi}^0) ^{3/2} v_0^{1/3} \alpha_T^{3/2}
  \label{scalemain}
\end{equation}
where $v_0$ is the \rev{ volume of the particles} . The real surface tension is then $\gamma=\mb{T} v_0 ^{-1} \hat \gamma $.
The surface tension therefore increases as a power law of the ratio between the  temperatures of the two types of particles $\alpha_T$ (see Appendix \ref{scaling} for details).

\subsection{Kinetics of droplet growth}\label{sec4b}
We consider a mixture quenched in the two phase region which is therefore 
supersaturated. We focus on a spherical droplet of phase-$b$ with 
radius $R$, growing inside the background phase which has a composition close to 
phase-$a$. The volume fractions outside the droplet are \rev{not equal to} the concentrations in the $a$ phase due to the supersaturation, i.e., $\phi_{\mathcal{A}}\rightarrow 
\phi_{\mathcal{A}}^{a}+\delta \phi_{\mathcal{A}}$ and  $\phi_{\mathcal{B}}
\rightarrow \phi_{\mathcal{B}}^{a}+\delta \phi_{\mathcal{B}}$.  Note that the concentrations inside the droplet are also slightly different from the concentrations of the $b$ phase but we can ignore this difference here.

This is a multi-scale \rev{problem with} two well separated length scales: the width of the interface $\ell$ is much smaller than the droplet size $R$. At the scale of the small length $\ell$, the system is still close to equilibrium with chemical potentials $\hat{\mu}_{\mathcal{A}}'$ and $\hat{\mu}_{\mathcal{B}}'$ which are the chemical potentials slightly shifted from the phase equilibrium values and calculated outside the droplet on its surface.
The 
steady-state equations, which give the particle concentration profiles are 

\begin{align}
\frac{\partial \hat{f}}{\partial \phi_{\mathcal{A}}}-\hat{\mu}_{\mathcal{A}}'=&\hat{L}_{\mathcal{A}}\left(\frac{\partial^2 \phi_{\mathcal{A}}}{\partial r^2}+\frac{2}{r} \frac{\partial \phi_{\mathcal{A}}}{\partial r}\right)\nonumber\\&+\hat{L}_{\mathcal{AB}}\left(\frac{\partial^2 \phi_{\mathcal{B}}}{\partial r^2}+\frac{2}{r} \frac{\partial \phi_{\mathcal{B}}}{\partial r}\right),\\\nonumber\\
\frac{\partial \hat{f}}{\partial \phi_{\mathcal{B}}}-\hat{\mu}_{\mathcal{B}}'=&\hat{L}_{\mathcal{B}}\left(\frac{\partial^2 \phi_{\mathcal{B}}}{\partial r^2}+\frac{2}{r} \frac{\partial \phi_{\mathcal{B}}}{\partial r}\right)\nonumber\\&+\hat{L}_{\mathcal{AB}}\left(\frac{\partial^2 \phi_{\mathcal{A}}}{\partial r^2}+\frac{2}{r} \frac{\partial \phi_{\mathcal{A}}}{\partial r}\right).
\label{chemeq}
\end{align}

In order to solve the so-called inner problem at the length scale $\ell$, we 
choose a length $\delta$ such that $\ell \ll \delta \ll R$. We multiply the first 
equation by $\partial \phi_{\mathcal{A}}/\partial r$ and the second one by  
$\partial \phi_{\mathcal{B}}/\partial r$, add them up and then integrate 
across the interface  over a region of size $\delta $ such that in both 
phases, $\partial \phi_{\mathcal{A}}/\partial r=\partial \phi_{\mathcal{B}}/
\partial r=0$ away from  the interface. As a result, we obtain the Gibbs-Thomson 
relation \cite{langer1985} :
\begin{eqnarray}
\delta \hat{\mu}_{\mathcal{A}} \Delta \phi_{\mathcal{A}}^{ab}+\delta \hat{\mu}_{\mathcal{B}} \Delta \phi_{\mathcal{B}}^{ab}=\frac{2\hat{\gamma} }{R}
\end{eqnarray}
where for each species,  $\Delta \phi_{\alpha}^{ab}=\phi_{\alpha}^{b}-\phi_{\alpha}^{a}$ and $\delta \hat \mu_{\alpha}=\hat \mu_{\alpha}^{\dagger}-\hat \mu'_{\alpha}$.  Note that outside the droplet, as discussed below, the 
volume fractions and the chemical potentials vary over the large length scale $R$ and do not change over the length $\delta$. As in the previous section, we now transform the volume fractions to normal  coordinates and 
obtain:

\begin{eqnarray}\label{gbt2}
\delta \hat{\mu}_{\eta} \Delta \eta_{ab}+\delta \hat{\mu}_{\psi} \Delta \psi_{ab}=\frac{2\hat{\gamma}}{R}.
\end{eqnarray}
The non-critical variable $\eta$ is identical in the two phases so that $
\Delta \eta_{ab}= 0$. The small variation of the chemical potential $\delta \hat{\mu}_{\psi}$ can 
be obtained from Eq.\eqref{eqc7} as $\delta \hat \mu_{\psi} \simeq 
2k_{\psi}\psi_a^2 \delta \psi$ using Eq.\eqref{eqc10}  where $\delta \psi=
\psi(R) - \psi_{a}$ is the small shift of the order parameter on the surface of 
the droplet from its equilibrium value in phase-$a$. Then, rearranging Eq.\eqref{gbt2}, we have:

\begin{eqnarray}\label{drop1}
\delta \psi(R)=\frac{\hat{\gamma}}{R k_{\psi} \psi_{a}^2 \Delta \psi _{ab}}.
\label{surface}
\end{eqnarray}

We now study  the dynamics of the growing droplet by studying the outer 
problem, calculating the order parameter  profile of the droplet material. As 
the non-critical variable $\eta$ has the same value in the two phases and 
as we find numerically that its variation is very small, we will assume here 
that there is no flux associated to this variable. The problem then has a single conserved order parameter $\psi$. 

\begin{figure*}
  \centering \includegraphics[width=1.92\columnwidth]{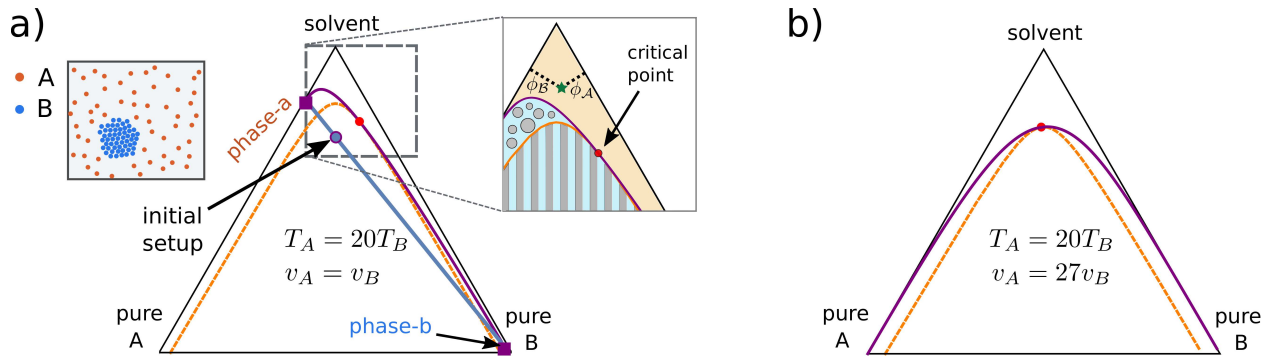}
  \caption{Triangular phase diagrams for three component system $
  \phi_{\mathcal{A}},\phi_{\mathcal{B}},\phi_s$(solvent) where $ \phi_{\mathcal{A}}+\phi_{\mathcal{B}}+\phi_s=1$; for every point in the 
  triangle the volume fractions are given by their distance from the facing 
  triangle side (exemplified on scaled inset by star). Both diagrams are for 
  hard-sphere systems ($a_{\alpha\beta}=0$, $\ma{\epsilon}=\mb{\epsilon}=8$) with 
  temperature ratio $\ma{T}/\mb{T}=20$ while size ratios differ. In a) $
  \ma{v}=\mb{v}$ whereas in b) $\ma{v}=27\mb{v}$. The purple and dashed 
  orange lines shows binodal and spinodal lines respectively and red 
  dots are the critical points. The location of the critical point is asymmetric in 
  both cases (though in opposite directions). Moreover, in a) we mark 
  two coexisting phases $a$ and $b$ (purple squares) which are 
  strongly asymmetric. Starting from the initial well-mixed setup, the system  phase 
  separates into a solid-like close-packed $\mathcal{B}$ particles (phase-$b
  $) surrounded by an $\mathcal{A}$ gas (phase-$
  a$) as illustrated in top left inset. In b) the phase diagram appears to be 
  more symmetric indicating both liquid-like coexisting phases 
  though shifted slightly towards $\mathcal{A}$-rich side. This results from having 
  $\alpha_T/\alpha_v=20/27\lesssim 1$, the ratio which controls 
  the symmetry of two phases (see Section \ref{sec3d})  }\label{fig2}
\end{figure*}

Outside the droplet, the order parameter $\psi$ follows a diffusion equation
$\partial \psi /\partial t = D_{\psi} \nabla^2 \psi $.
We discuss the value of the effective diffusion constant $D_{\psi}$ in Appendix \ref{dpsi}. The boundary conditions for this diffusion equation are the value $\delta\psi(R)$ given by the Eq.\eqref{surface} and the value at 
infinity $\psi_{\infty}$ which measures the supersaturation.
The solution of the diffusion equation is 
\begin{equation}
\psi(r)=\psi_{\infty}-(R/r) \left(\delta \psi_{\infty}-\delta \psi(R)\right).
\end{equation}
The growth of the droplet is due to the radial flux $j_{\psi}=-D_{\psi} \frac{\partial \psi}{\partial r}\vert_{r=R}$
The conservation of the flux of the order parameter $\psi$ on the surface of the droplet leads to $\Delta \psi_{ab} \frac{dR}{dt}=D_{\psi}\frac{\partial \psi}{\partial r} \vert_{R}$.
Inserting the solution of the diffusion equation, we obtain the evolution of the radius of the droplet
\begin{equation}
\frac{dR}{dt}=\frac{D_{\psi}}{R} \left(\Delta- \frac{d_0}{R}\right)
\end{equation}
where the supersaturation is defined as $\Delta=\delta \psi(\infty)/ \Delta 
\psi_{ab}$ \rev{and} $d_0= \hat \gamma / (k_{\psi}\psi_a^2\Delta\psi_{ab}^2)$ is a length of the order of the interfacial width $\ell$. This 
gives the critical nucleation radius of the droplet $R_c=d_0/\Delta$. 
Droplets smaller than $d_0$ collapse whereas droplets larger
than $d_0$ grow.

At the early stages of the phase separation just after the quench, there are 
few droplets and the droplets that are larger than the critical radius 
grow as $R\sim t^{1/2}
$. At long times, the value of the supersaturation decreases with time and a 
much more detailed analysis is required, which has been made by Lifshitz 
and Slyozov \cite{lifshitz1961}.  The supersaturation decreases as $\Delta \sim 
d_0/R$ and  the average droplet radius increases as $R\sim t^{1/3}$. Plugging in the value of $D_{\psi}$ obtained in Appendix \ref{dpsi} gives the scaling of droplet growth with time, $R\sim ( r_G t)^{1/3}$ in which the growth rate of mean droplet volume $r_G\sim (\phi_{\mathcal{A}}^0\alpha_T)^{1/2} v_0 ^{1/3}\mb{T} /\zeta$ that is linearly proportional to the geometric mean of $\ma{T}$ and $\mb{T}$.

\rev{Here we obtain these power laws in the dilute limit of our two-temperature model (which can be mapped on an equilibrium system)}. On the 
other hand, at higher order \rev{in the density expansion} even though the solution 
parameters ($\Delta, d_0, D_{\psi},r_G$) change, there is no reason to expect a different 
power law behavior than $R\sim t^{1/3}$ as long as the droplet material is transported by 
diffusion. Similar examples include one component active fluids which 
respect the  $1/3$ law \cite{wittkowski2014,lee2017}. 

\section{Some applications of the theory}\label{sec5}
\rev{In this 
section,  we show a few \rev{examples of application of the theory, to various} 
systems. Some of these results are in accordance with existing numerical studies while some may motivate future experiments and simulations.} 

As discussed earlier, the intra- and inter-species interactions can be controlled 
either by modifying the interaction potentials between particles 
or by changing the activity difference (the temperature ratio). In the first case 
one alters the virial coefficients while in the latter case, one 
changes both the entropic part of the effective free energy by changing the 
temperatures  $T_{\alpha}$ but also the  weight of the interactions through the 
pairwise temperatures $T_{\alpha\beta}$ as seen in Eq.~\eqref{fb}.
In general, for short-range interactions, the second virial coefficients can be 
written as $B_{\alpha\beta}\approx b_{\alpha\beta}+a_{\alpha\beta}/T_{\alpha\beta}$, where $b_{\alpha\beta}
$ is the effective pair excluded volume and $a_{\alpha\beta}$ is the additional 
interaction part. Moreover, $b_{\alpha\beta}$ can be approximated by its hard-core 
value. As an example, for spherical particles, $b_{\alpha\beta}\approx (4\pi/3)d_{\alpha\beta}
^3$ and $a_{\alpha\beta}\approx4\pi\int_{d_{\alpha\beta}} ^{\infty} u_{\alpha\beta}(r)r^2 
dr$ where $d_{\alpha\beta}$ is the distance between the centers of the particles at 
contact for an $\alpha \beta$ pair. If the additional interaction is purely 
attractive, $a_{\alpha\beta}<0$ is negative whereas it is positive if the additional 
interaction is repulsive. For a mixture of particles with given hard core 
sizes, 
this part of the virial coefficient can be tuned by chemical 
modifications. In 
addition, the contrast in activity which can be adjusted by the energy input per 
constituent, provides an extra handle 
\cite{palacci2013,theurkauff2012} for controlling the phase separation.  
\subsection{Colloidal hard spheres with different temperatures}
Pure hard spheres interact with each other only through excluded volume 
interactions, which do not allow them to interpenetrate:  $a_{\alpha\beta}=0$ 
and  $B_{\alpha\beta}=b_{\alpha\beta}$. The dramatic influence of activity contrast towards 
demixing is clearly observed in mixtures of hard spheres with equal sizes 
and hence equal mobilities \cite{weber2016}. In Fig.\ref{fig2}a we show the phase diagram for equal-sized hard spheres $
\alpha_v=1,\ \alpha_{\zeta}=1$, then $\ma{\epsilon}=\mb{\epsilon}=\mb{\beta}
=8$ \rev{with temperature ratio $\alpha_T=20$ \footnote{\rev{Note that $\alpha_T>4$ is the updated demixing condition for isometric hard spheres consistent with the second order virial expansion. In Ref.\cite{grosberg2015}, the authors had obtained $\alpha_T>34$ where they chose $\ma{\epsilon}=\mb{\epsilon}=1$ for simplicity}}}. The volume fractions of 
the cold and hot particles are not equal at the critical point where the volume 
fraction of the cold particles is larger, indicating asymmetric phases. 
As exemplified in the phase diagram, starting 
from a mixture in the unstable region, the system phase separates into a solid-like close-packed $\mathcal{B}$ particles (phase-$\beta$, which is not 
quantitatively well described by our low density approximation) surrounded 
by an $\mathcal{A}$ gas (phase-$\alpha$) as illustrated by top left inset. 

\begin{figure*}
  \centering \includegraphics[width=1.92\columnwidth]{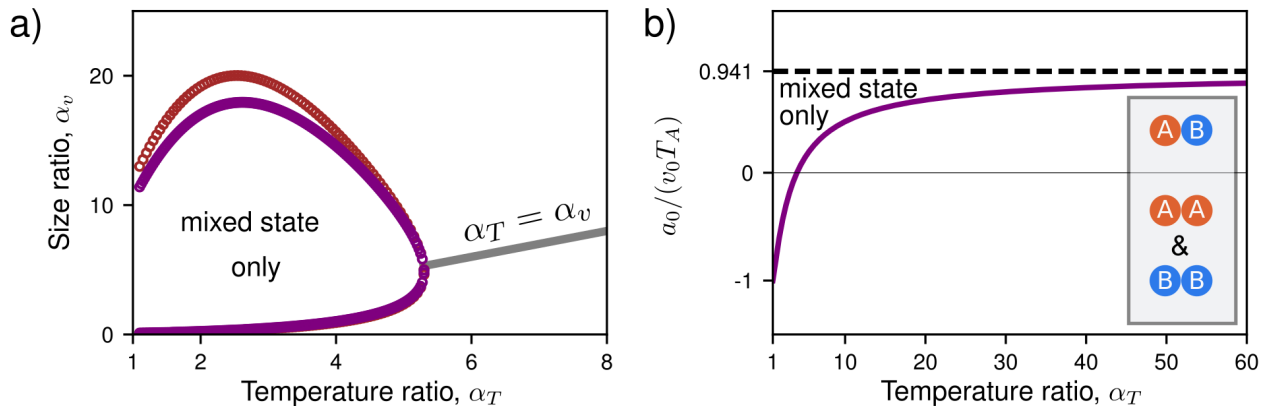}
  \caption{Control over mixing. a) Phase diagram for the existence of a phase separation \rev{in the} temperature ratio \rev{$\alpha_T=\ma{T}/\mb{T}$} and volume ratio \rev{$\alpha_v=\ma{v}/\mb{v}$ plane} for pure 
  hard spheres. The purple line follows the spinodal condition while the outer brown boundary more strictly requires the existence of a critical 
  point (see Section \ref{sec3e}). Inside these lines the solution remains mixed at all compositions. Interestingly, the phase diagram shows reentrance in 
  both the lateral and vertical directions. \rev{The gray line $\alpha_T=\alpha_v$ separates the different regimes of phase compositions: i) on lower half-plane, we expect solid-like $\mathcal{B}$-rich phase and gas-like $\mathcal{A}$-rich phase, and ii) vice versa on upper half-plane, iii) whereas near the gray line we expect both liquid-like phases.} In b) we use the same condition (in purple) for the existence of demixing for equal size hard spheres 
  with short-range interactions of equal magnitude but opposing behavior (attractive vs. repulsive) for intra- and inter-species. The inset shows the favored interactions in both regimes. If $a_0>0$, the 
  molecular interactions only promote mixing while at sufficiently large temperature ratios, the system can still reach  phase separation.  }
  \label{fig3}
\end{figure*}

Using our formalism, we can also investigate the phase behavior of hard-sphere mixtures with different  sizes. As we mentioned earlier, 
exclusively for active systems where $\ma{T}\neq \mb{T}$, do the mobilities (or 
the friction coefficients $\zeta_{\alpha}$) come into play via the 
effective pairwise temperature $T_{\mathcal{AB}}$ when $\ma{\zeta}\neq
\mb{\zeta}$. For a given size ratio $\alpha_v$, $\mb{\beta}=(1+
\alpha_v^{1/3})^{3}$ and using Stoke's law $\alpha_{\zeta}=\alpha_v^{1/3}$. In 
Fig.\ref{fig2}b we plot the phase diagram for mixtures of hard-spheres 
where the hot particles $\mathcal {A}$ are larger
with a volume ratio $\alpha_v=27$ and a temperature ratio $\alpha_T=20$. The evolution of the phase diagram upon increasing the size ratio 
can be appreciated by comparison to equal-size hard-spheres at the same temperature ratio (Fig.\ref{fig2}a-b). A more symmetrical phase diagram is 
predicted by increasing the volume ratio to reach $\alpha_v \approx \alpha_T$ and we expect coexistence between two liquid-like phases (Fig.\ref{fig2}b). A further increase of size ratio 
shifts the phase diagram towards the $\mathcal{A}$-rich side as expected for mixtures of hard spheres with different sizes at the same 
temperature. Another 
way to observe the effect of tuning both size and temperature ratios is to evaluate the conditions required for the existence of a phase separation 
given in Section \ref{sec3e} as a function of temperature and size ratios. We show this phase diagram in Fig. \ref{fig3}a which displays reenterances in both size ratio and temperature ratio axes.

\rev{\subsection{Relation to active swimmers and active-passive particle mixtures}\label{abp}}
\rev{In dilute mixtures} of active swimmers and passive particles, the run and 
tumble mechanism of the swimmers with propulsion speed $\ma{V}$ and 
reorientation time $\tau_r$ dictated by the time step between two tumbling events, can be considered at long times as entirely diffusive. If the 
reorientation time of swimmers is much smaller than the mean collision 
time \rev{(or mean free time)}, i.e., $\tau_r\ll \tau_c$, then $(\ma{T}-\mb{T})
\approx P_{cs} \tau_r/\rev{3}$ 
where $\mb{T}$ is the background temperature and 
$P_{cs}^{\mathcal{A}}=\ma{V}^2 \ma{\zeta}$ is the mean power required 
to drive chemotaxis \cite{mitchell1991} for an active 
particle. Accordingly, keeping $\tau_r$ high will serve lower dissipation rate to maintain the same \rev{translational} diffusivity. \rev{The validity range of this approximation in terms of the P\'{e}clet number $\text{Pe}=3 V_{\mathcal{A}}\tau_r/\ma{d}$ where $\ma{d}$ is the diameter can be obtained by estimating $\tau_c$ for active swimmers \cite{bruss2018phase} using collision theory. It suggests that the effective diffusive approximation $(\ma{T}-\mb{T})
\approx P_{cs} \tau_r/3$ remains valid for $\text{Pe}\ll \frac{3}{8\phi_{\mathcal{A}}}$ and density corrections are required as the concentration increases ($\tau_r\gg \tau_c$). Moreover, when we consider the emergence of the spinodal region as the demixing condition (Section \ref{sec3e}), we obtain at equal volume fractions $\phi_{\mathcal{A}}=\phi_{\mathcal{B}}=\phi_s$, the phase boundary follows $\phi_s\sim \text{Pe}^{-1}$. This relation is in accordance with simulations of mixtures of passive and active Brownian particles in two dimensions \cite{stenhammar2015activity}, although the simulations probe the phase separation at denser concentrations where both the diffusive approximation of the active swimmers fails (high Pe) and the dilute limit approximation of the  phase separation theory is not qualitatively accurate.}

\rev{One interesting aspect of our results is that we \rev{predict} a positive surface tension as discussed in the previous section. In single component active fluids, the contribution from the swim pressure \rev{ to the sign of surface tension is controversial. Certain approaches report negative \cite{bialke2015negative}, near zero \cite{omar2020microscopic} or positive \cite{hermann2019non} values, or even supporting both \cite{solon2018}}. For the two temperature model, at this order, the traditional "effective" equilibrium route coincides with the mechanical framework, however, we discuss the breakdown of this construction on Section \ref{sec6} while going one order higher in density. 

We can also express the cluster growth rate in terms of the mean power input. If the $\mathcal{A}$ and $\mathcal{B}$ constituents have the same volume $v_0$, at later stages of clustering, Eq.\eqref{eqe5} suggests that the growth rate of mean cluster size $ r_G \sim (\phi_{\mathcal{A}}P_{cs}\tau_r)^{1/2} v_0 ^{1/3} \mb{T} ^{1/2} /\zeta$ when $P_{cs}\gg \mb{T}/\tau_r$. This can be generalized similarly when there are two different types of swimmers and so on.} 

\subsection{Interacting particles with different temperatures}
Another interesting scenario appears when the interactions between particles enhance mixing  while the activity contrast opposes and boosts 
demixing. To illustrate  \rev{this scenario with} an example, consider a system of particles with equal strength of interactions which are repulsive for identical 
particles and attractive between different particles. This could be achieved for example by mixing hot and cold particles of opposite net electric charges in a medium with a finite screening length. Similar other systems can be prepared by \rev{engineering chemical} interactions. The interaction part of the virial coefficients can in this case be written as $a_{\mathcal{AA}}=a_{\mathcal{BB}}=a_0$ and 
$a_{\mathcal{AB}}=-a_0$. We may further simplify the problem by considering spherical particles of equal sizes such that $b_{\alpha\beta}=b_0=8v_0$ where $v_0$ is 
the molecular volume of the particles. We can evaluate the parameter range where a phase separation occurs. In Fig. \ref{fig3}b, we show the 
phase diagram in terms of the scaled interaction parameter $a_0/ (\ma{T} v_0)$, and the temperature ratio $\alpha_T=\ma{T}/\mb{T}$ for
$\ma{T}>\mb{T}$. When $a_0>0$, the 
  molecular interactions only promote mixing while at sufficiently large temperature ratios, the system can still reach  phase separation.

\subsection{Active-passive polymer blends}
Biopolymers play a key role in intra-cellular or intra-nuclear organization in many instances \cite{brangwynne2015}, while they often interact with active proteins which confer them an active character. In our context, such activity in biopolymers has been shown to enhance spatial segregation and maintain compaction. This is the case for example displayed in the structure and compaction of DNA inside the cell nucleus \cite{ganai2014,nuebler2018}. Similarly to colloidal particles, the active forces induce an effective temperature higher than the ambient one \cite{osmanovic2017}. 
To give an example within our theory, we consider here a mixture of poly-$\mathcal{A}$ and poly-$\mathcal{B}$ chains in solution with equal lengths but with different temperatures (or activities) $\ma{T}>\mb{T}$ and having only excluded-volume interactions. For dilute solutions, one can still expect an effective thermodynamic behavior with an effective free energy given by a Flory-Huggins theory. In the spirit of the Flory-Huggins mean-field theory, we suppose here that the interaction part of the free energy does not depend on the connectivity between the monomers and that we can employ the results obtained for colloidal particles. Therefore, in order to describe the polymers, we use the chemical potentials obtained in the section \ref{efftherm} by making the transformation $\mu_{\alpha}\rightarrow \mu_{\alpha}^{id}/\rev{N_{\alpha}} +\Phi_{\alpha}$, where \rev{$N_{\alpha}$} is the number of monomers in each chain.  Here, we consider the size interactions to be identical such that $\ma{\epsilon}=\mb{\epsilon}=\mb{\beta}=\epsilon_0$. By using the phase separation conditions described in Section \ref{sec3} (both approaches agree in this case when \rev{$N_{\alpha}\gg 1$}), we observe that segregation requires  \rev{$(\ma{T}-\mb{T})/\mb{T}=\alpha_T-1 >(2\epsilon_0^{-1/2})\left(\ma{N}^{-1/2}+\mb{N}^{-1/2} \right)$.} This result agrees with extensive simulations of active-passive polymer mixtures \cite{smrek2017} where the same scaling law is observed for \rev{$\ma{N}=\mb{N}=N$ as the condition becomes $\alpha_T-1 > (4\epsilon_0 ^{-1/2}) N^{-1/2}$}, though in terms of effective temperatures. \rev{Our mean-field exponents on the profile $\Delta \psi \sim \alpha_T ^{1/2}$, and interface width $\ell \sim \alpha_T ^{-1/2}$ seem close to the values obtained from simulations by the same authors in Ref.\cite{smrek2018}.}

\begin{figure*}
  \centering \includegraphics[width=1.85\columnwidth]{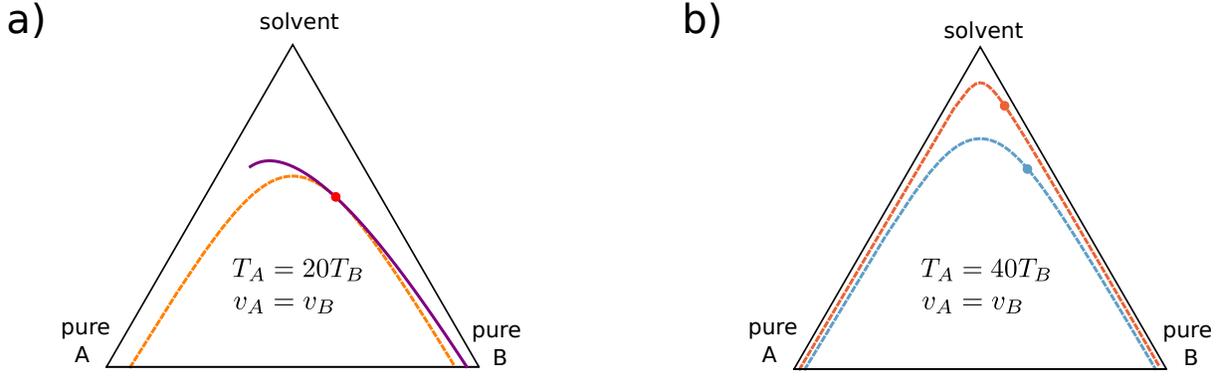}
  \caption{Evolution of triangular phase diagrams by third order contributions. 
  The presentation on how to read the components in the phase diagrams is 
  given in caption of \rev{Figure 2}. In a) we only show the resulting phase diagram 
  obtained using third-order expansion where we display binodal (purple), 
  spinodal (orange dashed) lines and the critical point (red) for equal-sized hard 
  spheres with $\ma{T}/\mb{T}=20$. In b) we compare the spinodal lines and 
  the critical points (for clarity, not the binodals) obtained from second-order 
  (orange) and third-order (light blue) expansions for equal-sized hard spheres 
  with $\ma{T}/\mb{T}=40$.}\label{fig4}
\end{figure*}
\rev{\section{Higher order expansions in concentration}\label{sec6}}
\subsection{General form}
Up to this point, we have studied the general phase behavior of a suspension of mixed 
particles with different temperatures in the dilute limit. In this limit,  the 
theory takes into account only two-particle correlations, the system has an 
effective thermodynamic description and the phase behavior 
can be obtained from 
the direct analog of 
the equilibrium construction of phase separation, despite the existence 
of non-equilibrium 
aspects such as the violation of  detailed balance that are observed at 
the microscopic level \cite{grosberg2015}.  A natural 
extension of this approach is then to calculate higher order corrections 
in concentration and see whether the effective thermodynamic description is 
preserved.

In order to answer this question, we must first obtain the steady-state pair 
distribution function $g_{\alpha\beta} ^{ss} ({\bf r}_1 - {\bf r}_2)$ at next orders in particle 
densities $c_{\alpha}$. A general strategy would be to start from the Fokker-Planck equation \eqref{FPmulti} for the multi-particle probability distribution $P
$, and integrate up
to the desired order in densities. One 
can then solve the remaining coupled equations to obtain the pair distribution 
functions $g_{\alpha\beta} ^{ss} ({\bf r}_1 - {\bf r}_2)$.  This approach leads to the 
Bogoliubov-Born-Green-Kirkwood-Yvon (BBGKY) hierarchy \cite{hansen1995}.  
For our problem, a non-equilibrium analog of this hierarchy is detailed in Ref. 
\cite{grosberg2015}. In the equilibrium case, the \rev{fluxes vanish and the distribution functions} are found by imposing a closure relation. By contrast, in a 
non-equilibrium case, for $\ma{T}\neq 
\mb{T}$, there might exist non-vanishing fluxes  associated to dissipation in the system \footnote{These fluxes are 
already present at the level of two-particle 
with different temperatures. As shown in Appendix A, the flux along the relative 
coordinate vanishes while a non-vanishing current exists along the center of 
friction coordinate.}. This complication makes it difficult to obtain a systematic 
expansion at higher orders in densities.  

A solvable example of Fokker-Planck equation (or the Langevin dynamics) at higher order has been given for pairwise harmonic potential interactions. In this case, a steady-state solution exists \cite{wang2019} 
for $\sum_m \partial_m 
J_m=0$ while the fluxes $J_m\neq 0$ when $\ma{T}\neq \mb{T}$. As a result, it is not possible to formulate a solution in Boltzmann form with a scalar temperature. 

Other classical approaches such as 
the Kirkwood superposition approximation would also fail. We refer the reader to the probabilistic interpretation of the Kirkwood closure in Ref.\cite{singer2004}.
 Nevertheless, in the following section, we demonstrate an alternative 
approach based on the calculation of depletion forces, to obtain third-order 
density corrections for pure hard-sphere interactions.

\subsection{Hard spheres}
In the case of mixtures with only hard-sphere interactions, an appropriate 
approach to expand at least to the next order, considers the  depletion 
interaction between two particles due to a third particle \cite{asakura1958}. This 
method has been used repeatedly in colloid science \cite{mao1995} and recovers exactly the 
third virial coefficients in hard-sphere mixtures with different radii. 
The full details of the calculation 
are given in Appendix \ref{appg}. Here as an example, we briefly sketch the results 
obtained for equal-sized hard-spheres  with different temperatures. 
The method is also 
applicable to particles of differing size ratios  as shown in the Appendix \ref{appg}. 
The resulting pressure is given by the standard virial expansion:

\begin{eqnarray}\label{virpress3rd}
p^0=\sum_{\alpha}T_{\alpha} c_{\alpha} + \rev{\frac{B}{2}}\sum_{\alpha,\beta} T_{\alpha\beta}  c_{\alpha} 
c_{\beta} + C\sum_{\alpha,\beta,\gamma} T_{\gamma}  c_{\gamma} c_{\alpha} c_{\beta}
\end{eqnarray}
where $B$ and $C$ are the second and third virial 
coefficients respectively, which are identical for all types of pairs and triplets of particles and $\alpha,\beta,\gamma=\mathcal{A}$ or $
\mathcal{B}$.

In equilibrium systems, there is a direct route from pressure to chemical potentials using the Gibbs-Duhem equation $p_0=\sum_{\alpha}\mu_{\alpha}^0 c_{\alpha}-f^0$, where the chemical potentials are given by 
$\mu_{\alpha}^0=\partial f^0/ \partial c_{\alpha}$. Our analysis for systems with two temperatures has shown that this remains applicable in the dilute limit approximation. However, without knowing explicitly 
the effective free energy, this approach is not founded. Thus, one should always start from the dynamic equations for the concentrations. 
If we insist in rewriting the interaction part of the chemical 
potentials defined in Eq.~\eqref{chempotinteract} as a series expansion in 
densities $\Phi_{\alpha}=\Phi_{\alpha} ^{(1)}+\Phi_{\alpha} ^{(2)}$, we only 
obtain the value of the chemical potential gradient at 
second order in the densities: 

\begin{equation}
\begin{split}
\nabla \Phi_{\mathcal{A}}^{(2)}=&C \bigg[\nabla \left(\frac{3}{2}\ma{T}\ma{c}^2+ 
\frac{3}{2}\mb{T}\mb{c}^2\right)\\&  +\left(2\mb{T}+\ma{T} \right)\mb{c}\nabla\ma{c}
+\left(2\ma{T}+\mb{T} \right)\ma{c}\nabla\mb{c}\bigg]
\end{split}\label{pressure3}
\end{equation}
where $C$ is the third virial coefficient between hard spheres of the same size 
(see Appendix \ref{appg}), which is equal for all types of interactions. At this order, $\nabla 
\Phi_{\mathcal{B}}^{(2)}=\nabla \Phi_{\mathcal{A}}^{(2)}$ because the mixture 
contains particles of  equal sizes. The total chemical potentials   
$\mu_{\alpha}^{(2)}=\mu_{\alpha}+ \Phi_{\alpha} ^{(2)} $ are then obtained from 
Eqs. \eqref{mueq}, \eqref{pressure3}. However, the gradient $\nabla \Phi_{\alpha} 
^{(2)}$ given by Eq.~\eqref{pressure3} is non-integrable to a potential 
function due to the mismatch between the cross terms. This incompatibility 
is related  to the 
non-equilibrium character of solutions of particles at different temperatures and hence the mismatch vanishes for an equilibrium solution, when $
\ma{T}=\mb{T}$. As a result, 
the non-equilibrium additional terms break down the routes to construct 
an effective 
thermodynamic theory. A similar breakdown has been previously observed in 
one component active fluids with density-dependent motilities 
\cite{wittkowski2014} where the interfacial contributions lead to 
terms not integrable 
 to a free energy. That issue has then been addressed in Ref. 
\cite{solon2018}, by introducing a functional transformation using an alternative 
scalar order parameter. By contrast, here we are not able 
to define even the local chemical 
potentials at third order in the dynamical equations ~\eqref{cform1}, \eqref{cform2}.  This could as well indicate the existence of bubbly phases 
\cite{tjhung2018} in which two steady state phases exist only locally and are 
separated by interfaces. A more detailed study would require to extend the 
analysis in gradient 
terms presented in the previous sections. We 
plan to address this question in a further study.

Since the term $\nabla \Phi_{\mathcal{A}}^{(2)}$ is not integrable to a chemical 
potential form, it is not straightforward to determine all uniformly conserved 
quantities at the zero-flux steady-state. At phase equilibrium, we only find two conserved quantities since $
\nabla p^0=0$ and $\nabla (\ma{\mu}^0-\mb{\mu}^0)=0$. The latter condition exists specifically
for equal-sized hard-spheres, because $\nabla 
\Phi_{\mathcal{B}}^{(2)}=\nabla \Phi_{\mathcal{A}}^{(2)}$. A third condition 
however, can be obtained by linearization of one of the concentration fluxes 
around the critical point. Then, the calculated phase diagrams are shown in Fig.\ref{fig4} in comparison to the phase diagram obtained from the previous 
expansion using second order virial coefficients (dilute limit) for the same set of 
parameters $\alpha_v=1$, $\alpha_T=40$. Curiously, the contribution from the third order 
terms delays the onset of demixing. \rev{We also observe that the instability occurs in physical region of the phase diagram for temperature ratios $\alpha_T>6.171$, not very far from the value obtained using second order analysis, i.e., $\alpha_T>4$ \cite{Note2}. }

\section{Conclusion}
To summarize, in this work, we have outlined the general framework to study the characteristics of phase equilibria in active mixtures of diffusive type where the temperatures of the constituents are different. We obtain the 
phase diagrams and phase growth properties using two methods in parallel which cover both the equilibrium thermodynamic description through functional analysis and a more general approach considering steady-state solutions 
of the Fokker-Planck equations. For each observable and method, we compare and contrast the non-equilibrium $\ma{T}\neq \mb{T}$ scenarios to the equilibrium ones $\ma{T}=\mb{T}$. This allows us to reexplore landmark 
methods and concepts developed to establish the foundational principles of equilibrium solution theory. Resemblances appear in the steady-state solutions while the non-equilibrium state when $\ma{T}\neq\mb{T}$ is only maintained with a net power dissipation. Even though these systems have a non-equilibrium character, it turns out that they recover a direct analog of the equilibrium construction in the dilute limit. Within this approximation, the direct connection with the equilibrium thermodynamics is linked to the closure of the hierarchy of correlation functions,which imposes a vanishing flux along the coordinate of interactions (the separation vector) for all two-body clusters \rev{(while this is no longer true for higher order clusters)}. In this dilute limit, we were able to construct a Cahn-Hilliard theory generalized to two temperature mixtures. This also validates a reasonable ground for phenomenological studies considering active systems as a perturbation around an equivalent equilibrium dynamics. 

\rev{The} analog of the equilibrium picture provides a rich palette to explore various aspects of active/passive (or less active) mixtures. In this case, the activity differences between the particles, introduce another level of control on the phase separation properties.  As we demonstrate by examples, the theory has broad applications in diverse 
 \rev{physical} systems at different length scales. \rev{We have also introduced} a transformation to normal coordinates around the critical state where the phase dynamics can be described by a single critical order parameter. \rev{We obtain counterparts of mean-field exponents of profile parameter $\psi$ to express interfacial properties, while $\eta$ becomes a measure of normal distance from the critical point (Similar to the temperature direction in regular solutions \cite{cahn1958,bray2002}, but not exactly the same since $\eta\equiv\eta(z)$ is a slightly varying function of $z$ along the interface).} In this simplified formalism, we capture interesting scaling laws for interfacial properties, droplet growth dynamics, and for the phase segregation condition. \rev{We observe that the surface tension is always positive at the interface of two phases for binary mixtures of equal-sized hard spheres. Some of our results are in agreement 
with existing numerical  simulations (detailed in Section \ref{sec5}). } Our results also \rev{suggest} a means to calibrate the composition of coexisting phases (liquid-liquid vs. gas-like and solid-like) by controlling the ratio $\alpha_T/\alpha_v$ which could motivate 
experimental applications. 

Higher order corrections (though not available for the general case) break down the direct analogy with the equilibrium construction in the case of pure hard-core interactions. However, the results do not indicate significant qualitative differences with the equilibrium behavior, and hence  in general, the qualitative behavior of the system should be obtained from the simpler theory describing the dilute limit (Sections \ref{sec2}-\ref{sec4}) that gives an intuitive understanding of demixing in diffusive systems. On the other hand, the existence of non-local terms in the chemical potentials at the third order in a power expansion in densities might indicate the emergence of new phenomena. It would be interesting to study these cases further including inhomogeneous terms, in order to bridge the gap between microscopic and coarse-grained models \cite{solon2018,tjhung2018}.


\begin{acknowledgments}
We thank A.Y. Grosberg, A.S. Vishen, and A.P. Solon for interesting discussions and insightful comments. E.I. acknowledges the financial support from the LabEx CelTisPhyBio. 
\end{acknowledgments}
\appendix

\section{Solution of the two-particle Fokker-Planck equation}
By evaluating Eq.~\eqref{FPmulti} for only two-particles $\alpha$ and $\beta$ at positions ${\bf r}_{1}$ and ${\bf r}_{2}$, and introducing a pairwise potential which depends only on the distance between these particles, $u_{\alpha\beta}\equiv u_{\alpha\beta}\left(|{\bf r}_{1}-{\bf r}_{2}|\right)$ such that $\partial_{{\bf r}_{1}} u_{\alpha\beta} =-\partial_{{\bf r}_{2}} u_{\alpha\beta} $, we can derive the steady-state solution using separation of variables where $\alpha, \beta =\mathcal{A}$ or $\mathcal{B}$. Accordingly, we set the two-particle probability function $P_{\alpha\beta}\equiv G_{\alpha\beta}({\bf R}) g_{\alpha\beta}({\bf r})$ where ${\bf r}={\bf r}_2-{\bf r}_1$  is the separation vector and  ${\bf R}=\tau_{\alpha}{\bf r}_1+\tau_{\beta}{\bf r}_2$ is the center of motion with $\tau_{\alpha}=\zeta_{\alpha} T_{\beta} /(\zeta_{\alpha} T_{\beta}+\zeta_{\beta} T_{\alpha})$ such that the diffusions along ${\bf r}$ and ${\bf R}$ are statistically independent \cite{grosberg2015}. As a result, we obtain:
\begin{eqnarray}\label{FPrel}
\frac{\partial P_{\alpha\beta}}{\partial t}= -\partial_r J_r^{\alpha\beta} - \partial_R J_R^{\alpha\beta}
\end{eqnarray}
with flux components:
\begin{align}
\label{jrel}
J_r^{\alpha\beta}&=\begin{aligned}[t]
            &-\left(\frac{\zeta_{\alpha}+\zeta_{\beta}}{\zeta_{\alpha}\zeta_{\beta}}\right) \frac{\partial u_{\alpha\beta}}{\partial {\bf r}}P_{\alpha\beta}\\&-\left(\frac{\zeta_{\alpha} T_{\beta}+\zeta_{\beta} T_{\alpha}}{\zeta_{\alpha}\zeta_{\beta}}\right)  \frac{\partial P_{\alpha\beta}}{\partial {\bf r}}\end{aligned}\\\nonumber\\
J_R^{\alpha\beta}&=\begin{aligned}[t]&-\left(\frac{T_{\alpha}-T_{\beta}}{\zeta_{\alpha} T_{\beta}+\zeta_{\beta} T_{\alpha}}\right) \frac{\partial u_{\alpha\beta}}{\partial {\bf r}}P_{\alpha\beta}\\&-\left(\frac{T_{\alpha} T_{\beta}}{\zeta_{\alpha} T_{\beta}+\zeta_{\beta} T_{\alpha}}\right)  \frac{\partial P_{\alpha\beta}}{\partial {\bf R}}.\end{aligned}
\end{align}
Setting the flux $J_r=0$ for the steady-state solution results Eq.\eqref{eqgss} in the main text. The remaining part only requires a uniform $G({\bf R})$ which satisfies the steady-state solution though $J_R^{\alpha\beta}$ does not necessarily vanish for $T_{\alpha} \neq T_{\beta}$ . 

\section{Irving-Kirkwood method for calculation of internal stress}\label{appb}
Following the stress equation for the interaction part, Eq.\eqref{ikstress} in the main text up to second order in separation vector ${\bf r}$ while using Eq.\eqref{eqgss} and keeping only non-vanishing terms upon integration, we rewrite the internal stress $\sigma_{ij} ^{(v)}$ given by Eq.~\eqref{irving} as:
\begin{equation}\label{ikstress2}
\begin{split}
\sigma_{ij} ^{(v)} = -p_0 \delta_{ij}  +\sum_{\alpha,\beta}&\frac{T_{\alpha\beta}}{2}\int \bigg[\frac{r_i r_j}{{\bf r}}  \frac{\partial }{\partial 
r}  \left(1-e^{-u_{\alpha\beta}(r)/T_{\alpha\beta}}\right) \\&  \bigg( \frac{1}{6}({\bf r}\cdot \nabla)^{2} c_{\alpha}({\bf r}_1) c_{\beta}({\bf r}_1)\\& -\frac{1}{2}[{\bf r}\cdot \nabla c_{\alpha}({\bf r}_1)] [{\bf r}\cdot \nabla  c_{\beta}({\bf r}_1) ] \bigg)  \bigg] d {\bf r},
\end{split}
\end{equation}
where $p^0=\left(\mu^0_{\mathcal{A}}c_{\mathcal{A}}+\mu^0_{\mathcal{B}}c_{\mathcal{B}}-f^0\right)$ is the locally uniform pressure. Integrating by parts gives:
\begin{equation}\label{ikstress3}
\begin{split}
\sigma_{ij} ^{(v)} = -p_0 \delta_{ij}  +\sum_{\alpha,\beta}\frac{T_{\alpha\beta}}{2}\mathcal{I}_{ijkl} &\bigg[\frac{1}{12}\left(\partial_k c_{\alpha}  \partial_l c_{\beta}+ \partial_k  c_{\beta}  \partial_l c_{\alpha}\right)\\&-\frac{1}{6}\left(c_{\alpha} \partial_k  \partial_l c_{\beta}+c_{\beta} \partial_k  \partial_l c_{\alpha} \right)\bigg]
\end{split}
\end{equation}
where a summation on the $k,l$ indices is performed using an Einstein summation convention, and the integral $\mathcal{I}_{ijkl}$ is given by:
\begin{equation}
\mathcal{I}_{ijkl}=\int \frac{r_i r_j r_k r_l}{r^2} \left(1-e^{-u_{\alpha\beta}(r)/T_{\alpha\beta}}\right)  d{\bf r}.
\end{equation} 
Considering the symmetries and performing the sum, we finally reach an expression for the difference between stress obtained by two different methods:
\begin{equation}\label{ikstress4}
\begin{split}
\sigma_{ij} ^{(v)}-\sigma_{ij}=\sum_{\alpha,\beta}L_{\alpha\beta}\bigg[-\frac{1}{3}\nabla^2 \left(c_{\alpha}  c_{\beta} \right)\delta_{ij}+\frac{1}{3}\partial_i \partial_j \left( c_{\alpha}  c_{\beta}\right)\bigg]
\end{split}
\end{equation}
where $\sigma_{ij}$ is the result obtained by free energy deformation given in Eq.~\eqref{stresseq}.
Summing over $\alpha,\beta$ suggests that the addition of the gauge term discussed in Section \ref{sec2b}, i.e., $-\frac{1}{3}\nabla^2 \left[  L_{\mathcal{A}} c_{\mathcal{A}}^2 + 2 L_{\mathcal{AB}}  c_{\mathcal{A}} 
c_{\mathcal{B}} + L_{\mathcal{B}} c_{\mathcal{B}}^2 \right]$  to the original free energy $f$, \rev{leads back to the} Irving-Kirkwood formula.

\section{Transformation of order parameters to normal coordinates around the critical point}\label{normal}
Close to the critical point  $\{\ma{\phi}^*, \mb{\phi}^* \}$, the effective thermodynamic description becomes 
simpler if instead of using the volume fractions measured with 
respect to the volume fractions at the critical point $\delta\phi_{\alpha}= 
\phi_{\alpha}-\phi_{\alpha}^{*}$ as variables, we make a linear transformation to the 
eigenvectors of the inverse compressibility matrix $\kappa^{-1}_{p}$ at the 
critical point. The inverse compressibility matrix is a symmetric matrix given by Eq. \eqref{fluceq} and it can be written as 
 \begin{eqnarray}
\kappa^{-1}_{p} =\left(
 \begin{matrix}
 c+a &   b \\
 b &   c-a
 \end{matrix}\right)
 \label{kappa_p}
 \end{eqnarray}
 The values of the coefficients are given by Eq.\eqref{fluceq} $c=\frac 1 2 \left[\alpha_T \left(\frac{1+\ma{\epsilon}\ma{\phi}}{\ma{\phi}} \right)+
 \alpha_v \left(\frac{1+\mb{\epsilon}\mb{\phi}}{\mb{\phi}} \right) \right]$, $a =
  \frac 1 2 \left[\alpha_T \left(\frac{1+\ma{\epsilon}\ma{\phi}}{\ma{\phi}} \right)-
 \alpha_v \left(\frac{1+\mb{\epsilon}\mb{\phi}}{\mb{\phi}} \right) \right]$ and $b=\frac{\alpha_{T}+\alpha_{\zeta}}{1+\alpha_{\zeta}}\mb{\beta}$. We denote the two eigenvalues of the matrix by $\epsilon$ and $\lambda$. 
 The coefficients of the matrix are real and related to the eigenvalues by 
 $c=\frac{\lambda + \epsilon}{2}$ and $(a^2 + b^2)^{1/2}=\frac{\lambda -
 \epsilon}{2}$. In the vicinity of the critical point, $\epsilon$ is small and
vanishes as one approaches the critical point and the other eigenvalue $\lambda$ remains finite at the critical point. The eigenvector associated with $\epsilon=0$ at the critical point, gives the direction in which the fluctuations diverge. We also define an angle $\theta$ such that $a=-
 \frac{\lambda -\epsilon}{2} \cos 2\theta$ and $b=\frac{\lambda -\epsilon}{2} \sin 
 2\theta$. The diagonalization appears then as a rotation of angle $\theta$ and the eigenvalue matrix $D_{\epsilon,\lambda}$ with diagonal entries $\epsilon$ and $\lambda$ is obtained by  rotation to the basis of eigenvectors $D_{\epsilon,\lambda}=R(\theta)\kappa^{-1}_{p} R^T(\theta)$ using the standard rotation matrix $R(\theta)$.
 
 In the volume fraction space, the eigenvectors are obtained using a 
 rotation of the natural coordinates by an angle $\theta$. The normal coordinates that we use are the 
 coordinates along the eigenvectors of the inverse compressibility matrix at the critical point. At the critical 
 point, the rotation angle of the eigenvectors is $\theta^*$ that satisfies
 \begin{equation}
 \tan \theta^* =\frac{ b^*}{c^*-a^*}=  \frac{(\alpha_T+\alpha_{\zeta})\mb{\beta}\mb{\phi}^*}{\alpha_v (1+
 \alpha_{\zeta})(1+\mb{\epsilon}\mb{\phi}^*)}
 \label{thetac}
 \end{equation}
 which can also be expressed in conjugate form in terms of $\ma{\phi}^*$ using $\tan \theta^* =\frac{ c^*+a^*}{b^*}$. This angle gives the orientation of the tie lines close to the critical point. As a result, it provides an indication on the asymmetry of composition between the two phases.  Thus, when $\tan\theta^*\approx 1$ , we would have two liquid-like phases, whereas for $\tan \theta^*\gg 1$, the critical point is towards $\mathcal{B}$-rich side of the phase diagram with a solid-like and a gas-like phase coexisting and vice versa for $\tan \theta^*\ll 1$.
 
 In the eigenbasis of the inverse compressibility matrix at the critical point, there is a coordinate $ \psi$ 
 along the eigenvector associated to the vanishing eigenvalue and a coordinate $\eta$ along the 
 eigenvector associated to the finite eigenvalue $\lambda$. These two normal coordinates are related to 
 the original volume fractions by the rotation matrix $R(\theta^*)$. Accordingly, we define:
\begin{equation}
\left(
 \begin{matrix} 
\psi \\
 \eta
 \end{matrix}\right)=R(\theta^*) \left(
 \begin{matrix}
\delta \ma{\phi} \\
\delta \mb{\phi} 
 \end{matrix}\right).
\end{equation} 
The coordinate $\psi$ is the critical variable that we call the order parameter. Along these new coordinates, differentiation is performed as:
\begin{equation}\label{eqc4}
\left(
 \begin{matrix}
\frac{\partial}{\partial\psi} \\
 \frac{\partial}{\partial\eta}
 \end{matrix}\right)=R(\theta^*) \left(
 \begin{matrix}
\frac{\partial}{\partial \ma{\phi}} \\
 \frac{\partial}{\partial \mb{\phi}}
 \end{matrix}\right). 
\end{equation} 
Using these relations, we expand the free energy around the critical point and obtain:

\begin{equation}\label{eqc5}
\hat f_0= \hat f_0^* + \hat \mu_{\eta}^* \eta+ \hat \mu_{\psi}^* \psi+ \frac { k_2} 2 \psi^2 \eta + \frac { k_4} 4 \psi^4 +\frac{\lambda^*} 2 \eta^2,
\end{equation}
where the coefficients of the expansion are the derivatives of the free energy $\hat f_0$ evaluated at the 
critical point: $k_2=\frac{\partial^3 \hat f_0}{\partial \psi^2 \partial \eta} \Big|_*$, $k_4=\frac{1}{3!}
\frac{\partial^4 \hat f_0}{\partial \psi^4} \Big|_*$, $\lambda^*=\frac{\partial^2 \hat f_0}{\partial \eta^2} \Big|
_*$ and we used the fact that $\eta$ is a slowly changing variable. The Hessian matrix of the second 
derivatives of the free energy is equal to the inverse compressibility matrix. In the coordinates $\psi, \eta$ 
the inverse compressibility matrix is $D_{\epsilon,\lambda}$ evaluated at the critical point where $
\epsilon=0$. The two derivatives $\frac{\partial^2 \hat f_0}{\partial \psi \partial \eta} \Big|_*$ and  $
\frac{\partial^2 \hat f_0}{\partial \psi^2} \Big|_*$ therefore vanish. The third derivative of the free energy $
\frac{\partial^3 \hat f_0}{\partial \psi ^3} \Big|_*$ also vanishes  because  $\psi$ is the tangent direction to 
the spinodal line at the critical point. The two  chemical potentials along the new coordinates are obtained 
again by differentiation of the free energy $\hat f_0$:
\begin{eqnarray}
  \hat \mu_{\eta} &=& \hat \mu_{\eta}^*+ \lambda^* \eta  + k_2 \frac{\psi^2}2\\
  \hat \mu_{\psi}&=&  \hat \mu_{\psi}^* +k_2 \eta \psi +{k_4} \psi^3  \label{eqc7}
 \end{eqnarray}
The coordinates of the equilibrium phases  $a$ and $b$ (the binodal line)  are 
obtained by equating the chemical potentials $\hat \mu_{\psi}$ and $\hat \mu_{\eta}$ 
and the pressure in the two phases. This leads to  $\eta_a=\eta_b$ and $\psi_a^2=\psi_b^2 =-\frac{k_2}{ k_4} \eta_a$.
By construction of the normal coordinates, the tie lines, which are the straight 
lines joining the two phases at equilibrium correspond to lines of constant 
values of $\eta$.  The binodal line has therefore a parabolic shape with $\psi_a\approx -\psi_b$. Each 
value of $\eta_a\approx \eta_b >0$ defines the coexisting phases $\psi_a$ and $\psi_b$. It is convenient in the following to 
consider a symmetrized version of the order parameter $\Delta \psi_{ab}=\psi_b-\psi_a$, which 
satisfies $\Delta \psi_{ab}^2=-\frac{4 k_2}{k_4}\eta_a$.

The total free energy density is obtained by including the gradient terms which also can be transformed to the normal coordinates. In the vicinity of the 
critical point, as the non-critical variable $\eta$ is much smaller than the 
order parameter $\psi$ we only need to retain terms in $({\bf \nabla} \psi)^2$. The total free energy density reads then 
\begin{equation}
 \hat f =\hat f_0 + \frac {\hat L_{\psi}}2 ({\bf \nabla} \psi)^2
 \label{totalf0}
 \end{equation}
 where the coefficient $\hat L_{\psi}$ is given by 
 $\hat L_{\psi} ={\hat L}_{\mathcal{A}} \cos^2 \theta^*-2{\hat L}_{\mathcal{AB}} \cos\theta^* \sin\theta^*+{\hat L}_{\mathcal{B}} \sin^2 \theta^*$.
 
 In order to calculate the interfacial tension, we must define the two phases 
 at equilibrium i.e. fix the values $\eta_a$ in the two phases in equilibrium. This also fixes the order parameter $\Delta 
 \psi_{ab}$. We must then calculate what we called the tilted free energy around the critical point  
 \begin{equation}
  \Delta \hat f [\psi, \eta]=\hat  f [\psi, \eta] - \hat f^0 (\psi_a, \eta_a)-\hat \mu_{\psi}^{\dagger} \left(\psi-
  \psi_a \right)- \hat \mu_{\eta}^{\dagger}\left(\eta-\eta_a \right) 
 \end{equation}
The constants $\hat \mu_{\psi}^{\dagger}$ and $\hat \mu_{\eta}^{\dagger}$ are the chemical potentials  
calculated in the two phases at equilibrium. Note that the tilted free energy has a minimum and vanishes 
in the two phases in equilibrium. The profiles of $\psi$ and $\eta$ as a function of the coordinate $z$ 
perpendicular to the interface are obtained by minimization of the tilted free energy.
We first minimize the tilted free energy with respect to $\eta$. This leads to 
 \begin{equation}\label{eqc10}
  \eta=\eta_a -\frac{k_2}{2\lambda^*}(\psi^2-{\psi_{a}}^2).
 \end{equation}
 Inserting this result into the tilted free energy, we obtain the tilted free energy as a function of the order parameter $\psi$ only, which we write as
 \begin{equation}
  \Delta \hat f (\psi)= \frac{k_{\psi}}4 (\psi - \psi_a )^2 (\psi - \psi_b )^2+\frac {\hat{L}_{\psi}}2 ({\bf \nabla} \psi)^2
  \label{totalf}
 \end{equation}
 where $k_{\psi}=  k_4 - \frac{k_2^2}{2\lambda^*}$. This free energy is minimal and vanishes at $\psi=\psi_{a}$ and $\psi=\psi_b$. Finally, minimization with respect to $\psi$ \rev{yields}:
  \begin{equation}
  \Delta \hat f^0 (\psi)= \frac{k_{\psi}}4 (\psi - \psi_a )^2 (\psi - \psi_b )^2=\frac {\hat{L}_{\psi}}2 ({\bf \nabla} \psi)^2
  \label{totalfzero}
 \end{equation}
 which shows that along the order parameter profile between the two phases, $\Delta \hat f (\psi)=2 \Delta 
 \hat f^0 (\psi)$.
 
 \rev{\section{Positivity of the surface tension}\label{appd}
 
 As mentioned in the main text, since $\hat{\gamma}= L_{\psi}\int (\nabla\psi)^2$ integrated from one phase to the other, the sign of $L_{\psi}$ determines the sign of the surface tension. We can determine the sign of $L_{\psi}$ using our results in Section \ref{sec2}.
 From the definition $\hat L_{\psi} ={\hat L}_{\mathcal{A}} \cos^2 \theta^*-2{\hat L}_{\mathcal{AB}} \cos\theta^* \sin\theta^*+{\hat L}_{\mathcal{B}} \sin^2 \theta^*$, the positivity of surface tension requires:
\begin{equation}
 2\tan \theta^*-\frac{{\hat L}_{\mathcal{A}}}{{\hat L}_\mathcal{AB}}-\frac{{\hat L}_{\mathcal{B}}}{{\hat L}_\mathcal{AB}}\tan^2 \theta^*>0.
\end{equation}
We have previously defined rescaled parameters $\hat{L}_{\alpha\beta}= 
\mb{T}^{-1}L_{\alpha\beta}\ma{v}/(v_{\alpha}v_{\beta})$ on Section III.A (then $\hat{L}_{\mathcal{A}}/\hat{L}_{\mathcal{AB}}=(L_{\mathcal{A}}/L_{\mathcal{AB}})\alpha_v ^{-1}$ and  $\hat{L}_{\mathcal{B}}/\hat{L}_{\mathcal{AB}}=(L_{\mathcal{B}}/L_{\mathcal{AB}})\alpha_v $ ) whereas $L_{\alpha\beta}=-T_{\alpha\beta}\Lambda_{\alpha\beta}$ and the coefficients $\Lambda_{\alpha\beta}=\frac{1}{6}\int  r^2 
(1-e^{-u^{\alpha\beta}({\bf r})/T_{\alpha\beta}}) d {\bf r}=\tilde{\sigma}_D B_{\alpha\beta}^{5/3}$ with $\tilde{\sigma}_D=\frac{1}{10}(3/4\pi)^{2/3}$ for hard spheres. Bringing these together and using $\ma{\epsilon}=\mb{\epsilon}=8$ suggests the positivity condition for the surface tension to be:
\begin{equation}
 2\alpha_v\tan \theta^*-\frac{1+\alpha_{\zeta}}{\alpha_T+\alpha_{\zeta}}\left(\frac{8}{\beta_{\mathcal{B}}}\right)^{5/3}\left( \alpha_T\alpha_v^{5/3}+\alpha_v ^2 \tan^2 \theta^*\right)>0.
\end{equation}
We can then use the definition of $\tan \theta^*$ from Eq.\eqref{thetac}, and $\alpha_{\zeta}=\alpha_v ^{1/3}$, $\beta_{\mathcal{B}}=(1+\alpha_v ^{1/3})^3$. For \rev{the specific cases $\alpha_v=1$ and $\alpha_v=\alpha_T$, it is easy to prove the positivity of $L_{\psi}$}. For varying size ratios, \rev{a numerical evaluation} shows that $L_{\psi}$ is positive in the region where demixing occurs except when $\alpha_v \ll 1$ or $\alpha_v \gg \alpha_T$. In these two extreme limits, the surface tension can result a negative value, although it is not clear whether this is a true behavior since for extreme size ratios, the flat interface assumption would also fail. }

\section{Scaling relations for equal-sized hard spheres when $\alpha_T\gg 1$}\label{scaling}
As shown in the main text for equal-sized hard spheres where $\alpha_v=1$, (and hence $\alpha_{\zeta}=1$), and 
$\ma{\epsilon}=\mb{\epsilon}=\mb{\beta}=8$, the volume fractions at the critical point are $\ma{\phi}^*=\alpha_T ^{-1}$, 
$\mb{\phi}^*=1/8+(5/4)\alpha_T^{-1}$ when $\alpha_T \gg 1$.  Using these relations, the rotation angle at the critical point 
given by Eq. \eqref{thetac} reads,
\begin{equation}
\tan \theta^* \approx \frac{\alpha_T} 4 .
\end{equation}
 We consider a mixture with average particle volume fractions $\ma{\phi}^0$ and $\mb{\phi}^0$ 
 defined by $\phi_{\alpha}^0=V^{-1}\int_V \phi_{\alpha}({\bf r}) d{\bf r}$. We determine the 
 normal distance $\eta(\ma{\phi}^0,\mb{\phi}^0)$ from the critical point. Then, 
 as discussed in Appendix \ref{normal}, this fixes the order parameter $\Delta \psi_{ab}^2=-\frac{4 k_2}{k_4} 
 \eta(\ma{\phi}^0,\mb{\phi}^0)$, and hence the compositions of the coexisting phases. \rev{In the limit where $
 \alpha_T\gg 1$, the order parameter is $\Delta \psi_{ab}^2\approx \alpha_T (\ma{\phi}^0-\ma{\phi}^*)/4+(\mb{\phi}^0-\mb{\phi}^*)$ which becomes
\begin{equation}\label{eqd2}
\Delta \psi_{ab}^2\approx\frac{\ma{\phi}^0}{4}\alpha_T
\end{equation}
at (reasonably) finite volume fractions in the vicinity of the critical point.}

In order to discuss the scaling of the surface tension, we also need the expressions of the coefficients $k_\psi$ and 
$\hat{L}_{\alpha\beta}$ as functions of $\alpha_T$.  For equal-size hard spheres, we have:
\begin{eqnarray}
\ma{\hat{L}}=-\sigma_D\alpha_T v_0 ^{2/3}, \hspace{1em} \mb{\hat{L}}=-\sigma_D v_0 ^{2/3}, \nonumber\\ 
\hat{L}_{\mathcal{AB}}=-\sigma_D \frac{(1+\alpha_T)}{2} v_0 ^{2/3}
\end{eqnarray} 
where $\sigma_D=\frac 4 5 (6/\pi)^{2/3}$ and $v_0$ is the molecular volume. We can then use these results to obtain $\hat L_{\psi} 
={\hat L}_{\mathcal{A}} \cos^2 \theta^*-2{\hat L}_{\mathcal{AB}} \cos\theta^* \sin\theta^*+{\hat L}_{\mathcal{B}} \sin^2 
\theta^*$. The other coefficient $k_{\psi}=k_4-\frac{k_2^2}{2\lambda^*}$ can be calculated by taking into account the definitions below 
Eq.\eqref{eqc5} and  differentiating Eq.\eqref{eqc4}. For $\alpha_T\gg1$, $k_4\approx 256 \sin^4\theta^*$, 
$k_2\approx -16\alpha_T\sin^3\theta^* $ and $\lambda^*\approx \alpha_T ^2\sin^2\theta^*$. Accordingly, $k_{\psi}\approx 128$ in this limit which is independent of the temperature ratio similarly to $\hat{L}_{\psi}$. These relations using \eqref{gammasc1} lead to Eq.\eqref{scalemain} in the main text.

\section{Calculation of $D_{\psi}$ and the droplet growth rate}\label{dpsi}
In order to determine the diffusion coefficient of the relevant order parameter $\psi$ given in \ref{sec4b}, we need first to linearize time evolution equations \eqref{cform2} around one of the phases (say phase-$a$). This suggests:
\begin{equation}\label{diffeq1}
\frac{\partial}{\partial t}\left(
 \begin{matrix}
\delta \ma{\phi} \\
\delta \mb{\phi}
 \end{matrix}\right)=\Gamma^a \nabla^2 \left(
 \begin{matrix}
\delta \ma{\phi} \\
\delta \mb{\phi} 
 \end{matrix}\right).
\end{equation} 
 in which we define  $\delta \phi_{\alpha}=\phi_{\alpha}-\phi_{\alpha}^*$ and obtain $\Gamma^a$ by evaluating \eqref{gameq2} at phase-$a$. Then, following Appendix \ref{appb}, we obtain:
\begin{equation}
\frac{\partial}{\partial t}\left(
 \begin{matrix}
\psi \\
 \eta
 \end{matrix}\right)=\Gamma'^a \nabla^2\left(
 \begin{matrix}
\psi\\
 \eta
 \end{matrix}\right).
\end{equation} 
where $\Gamma'^a=R(\theta^*)\Gamma^a R^T(\theta^*)$. Using $\partial \eta /\partial t \approx 0$, we have $D_{\psi}=|\Gamma'^a|/\Gamma'^{a}_{22}$ with $|\Gamma'^a|=|\Gamma^a|$ being the determinant of matrix $\Gamma'^a$ and $\Gamma'^a_{22}$ is the second diagonal component. Around the critical point, using Eqs.\eqref{eqc4},\eqref{eqc7} and \eqref{eqc10}, the diffusion coefficient is:
\begin{equation}
D_{\psi}\approx \frac{\mb{T} \ma{\phi}^* \mb{\phi}^* }{\mb{\zeta}\alpha_v \left(\ma{\phi}^*\sin^2\theta^*+\mb{\phi}^*\cos^2\theta^* \alpha_v^{-1}\alpha_{\zeta} \right)} 2k_{\psi} \psi_a^2
\end{equation}
where $\ma{\phi}^a $ and $\ma{\phi}^a $ are the values of volume fractions at phase-$a$. At high temperature ratios $\alpha_T\gg 1$, we see that $\sin\theta^*\approx 1$ and hence:
\begin{equation}
D_{\psi}\approx \frac{ \mb{T}}{4 \zeta}k_{\psi} \psi_a^2
\end{equation}
where we consider equal-size hard-spheres, $\alpha_v=\alpha_{\zeta}=1$ and $\ma{\zeta}=\mb{\zeta}=\zeta$.

Next, we estimate the growth rate of average droplet size $r_G$ such that $R\sim (r_G t)^{1/3}$. Near saturation $\Delta\sim d_0/R$ and hence $r_G \sim D_{\psi} d_0$. Using the above relation and the value of $d_0$ from the main text, we obtain $r_G \sim \mb{T}\Delta \psi_{ab}v_0 ^{1/3}/\zeta$. Finally, since $\Delta \psi_{ab}\sim (\phi_{\mathcal{A}}^0\alpha_T)^{1/2}/2$ from \eqref{eqd2}, we have 
\begin{equation}\label{eqe5}
r_G \sim (\phi_{\mathcal{A}}^0\alpha_T)^{1/2}v_0 ^{1/3} \mb{T} /\zeta.
\end{equation}

\section{Third order expansion for pure hard-spheres}\label{appg}
\subsection{Virial expansion to third order}
The strategy that we follow in this appendix is to define a potential of mean force at a finite concentration 
$u_{\alpha\beta}^{tot}(r)$ for an  $\alpha \beta$ pair and to insert it into the Fokker-Planck equation \eqref{FPrel}, \eqref{jrel} by replacing $u_{\alpha\beta}\rightarrow u_{\alpha\beta}^{tot}$. This results in the steady-state value of the pair 
distribution function written in the Boltzmann form 
$g_{\alpha\beta}^{ss}(r)=\exp\left(-u_{\alpha\beta}^{tot}(r)/T_{\alpha\beta} \right)$
where $g_{\alpha\beta}$ is the pair distribution function, $u_{\alpha\beta}^{tot}(r)$ is the potential of mean force, 
and $T_{\alpha\beta}$ is the pairwise temperature between the two particles under consideration. In general, we can 
separate the potential of mean force into two parts, $u_{\alpha\beta}^{tot}(r)= u_{\alpha\beta}(r)+W_{\alpha\beta}(r)$ where 
$u_{\alpha\beta}(r)$ is the bare pair interaction potential for two particles while $W_{\alpha\beta}(r)$ \cite{kirkwood1935} 
 is due to the existence of other surrounding particles. The potential of mean force can be derived 
by geometrical considerations for hard-sphere mixtures. 

When the excluded volumes between the two particles overlap such that a third particle $\gamma$ cannot enter in 
the space between the two particles, there is a net attractive force between the two particles known as the depletion force \cite{asakura1958}. The 
mean force between an $\alpha \beta$ pair due to 
a third particle is given as $F_{\alpha\beta}=-p_0 ^{tot} S$ where $p_0 ^{tot}$ is the total pressure of third body 
particles and $S(r)$ is the cross-sectional area loss at the overlapping region, which depends 
on the distance $r$  
between the two particles of an $\alpha \beta$ pair. For equal-size particles with diameter $d$, 
this surface is given by:
\begin{eqnarray}
S(r)=
 \begin{cases} 
      \pi (d^2-r^2/4) & d\leq r \leq 2d \\
      0 & r\geq 2d
   \end{cases}
\end{eqnarray} 
The integration of this force $F_{\alpha\beta}$ yields the potential of mean force and the interaction $W_{\alpha\beta}(r)$ between 
the $\alpha \beta$ pair mediated by a third particle within the region $d\leq r\leq 2d$. By imposing continuity 
of the interaction potential, we 
obtain:
\begin{eqnarray}\label{uanw}
W_{\alpha\beta}(r)&=&-p_0 ^{tot} w(r)\\ \nonumber
w(r)&=&\frac{\pi}{12}\left(16d^3-12d^2r+r^3 \right)\Theta(2d-r)
\end{eqnarray}
where $w(r)$ is the overlap volume and $\Theta(x)$ is Heaviside step function. 
Note that since $p_0^{tot}$ 
is a function of the third particle only, the mean force and the corresponding potential are independent of 
the $\alpha\beta$ pair, i.e., $W_{\alpha\beta}(r)=W(r)$. In a more formal manner, this potential can be expressed as:
\begin{eqnarray}\label{umnfull}
W_{\alpha\beta}=-\sum_{\gamma} \int T_{\gamma} c_{\gamma} ({\bf r}_{\gamma}) \Theta (d-r_{\alpha\gamma}) \Theta (d-r_{\beta\gamma}) d {\bf r}_{\gamma}.\nonumber\\
\end{eqnarray}
Then by using the piecewise definition of the hard-sphere potential we write, up to first order in concentrations:
\begin{eqnarray}\label{gmnapprx}
g_{\alpha\beta}(r)\approx e^{-u_{\alpha\beta}(r)/T_{\alpha\beta}} \left[1-\frac{W_{\alpha\beta}(r)}{T_{\alpha\beta}} \right]
\end{eqnarray}
where the pairwise hard-sphere potential $u_{\alpha\beta}(r)=\infty$ for $r<d$ and $u_{\alpha\beta}(r)=0$ elsewhere.

We now use this pair distribution function to obtain thermodynamic properties. The pressure is obtained from the virial equation which becomes in the case of hard-spheres of equal-sizes:
\begin{eqnarray}
p=\sum_{\alpha} T_{\alpha} c_{\alpha} + \frac{2}{3} \pi d^3 \sum_{\alpha,\beta} c_{\alpha} c_{\beta} g_{\alpha\beta}(d) T_{\alpha\beta}.
\end{eqnarray}
Inserting the expression of the pair distribution function Eq.~\eqref{gmnapprx}, we obtain the pressure as:
\begin{eqnarray}\label{virpress}
p=\sum_{\alpha}T_{\alpha} c_{\alpha} + \frac{2}{3} \pi d^3 \sum_{\alpha,\beta}   c_{\alpha} c_{\beta}   \left[T_{\alpha\beta}+ w(d) \sum_{\gamma}c_{\gamma} T_{\gamma} \right]\nonumber \\
\end{eqnarray}
The second and third virial coefficients are therefore \rev{$B=\frac{4\pi}{3}d^3$} and $C=\frac{2\pi}{3}d^3 w(d)$ 
with  $w(d)=\frac{5\pi}{12}d^3$ from Eq.\eqref{uanw}. Rewriting in terms of the virial coefficients gives 
Eq.\eqref{virpress3rd} in main text. Note that the third order terms in the expansion are weighted by the 
temperature of the third particle. Thus, introducing different mobilities will not alter the form of this 
equation. 

The next task is to implement the same strategy to the time evolution equations for the concentrations 
$\ma{c}$ and $\mb{c}$, given by Eqs. \eqref{cform1}, \eqref{cform2}. In a similar way, we define 
the total mean force $\bar{f}_{\alpha}$ on particles $m$, as the gradient of a potential $\bar{f}_{\alpha}=-
\nabla \Phi_{\alpha}$.  Using Eq.\eqref{gmnapprx}, the interaction part of the chemical potentials is 
expanded in powers of concentrations as $\ma{\Phi}=\ma{\Phi}^{(1)}+\ma{\Phi}^{(2)}$. The lowest order 
term $\ma{\Phi}^{(1)}$ has been obtained in terms of the second virial coefficients in Eq.~\eqref{mueq}. We 
obtain the next order contribution using Eqs.\eqref{umnfull} and \eqref{gmnapprx}:

\begin{align}
\nabla_{{\bf r}_1}\Phi_{\alpha}^{(2)}&=\sum_{\beta,\gamma} \bigg[\int \frac{\partial}{\partial {\bf r}_1} \left(1-e^{-u_{\alpha\beta}(r)/T_{\alpha\beta}} \right) c_{\beta} ({\bf r}_2) \\&  \int T_{\gamma} c_{\gamma} ({\bf r}_3) \Theta (d-r_{13}) \Theta (d-r_{23}) d {\bf r}_3 d {\bf r}_2\bigg]\nonumber.
\end{align}
By coordinate transformation, we get:

\begin{align}\label{phieq}
\nabla_{{\bf r}_1}\Phi_{\alpha}^{(2)}&=\sum_{\beta,\gamma} \bigg[\int \widehat{{\bf r}} \hspace{2pt}\delta (r-d) c_{\beta} ({\bf r}_1 + {\bf r})\\& \nonumber \int T_{\gamma} c_{\gamma} ({\bf r}_1+{\bf r}') \Theta (d-r') \Theta (d-|{\bf r}- {\bf r}'|)  d {\bf r}' d {\bf r}\bigg]
\end{align}
with ${\bf r}={\bf r}_2-{\bf r}_1$ and ${\bf r'}={\bf r}_3-{\bf r}_1$ where $\widehat{{\bf x}}$ denotes the unit vector along the direction of a vector ${\bf x}$. The second 
integral is constrained over the overlap volume ${V_{\cap}}$. Taking into account the Dirac delta function 
$\delta (r-d)$, it can be expressed as:

\begin{align}
\int\limits_{{V_{\cap}}} T_{\gamma} c_{\gamma} ({\bf r}_1+{\bf r}') d {\bf r}' = \int\limits_{0} ^{2\pi} d\phi' \int\limits_{0} ^{d} dr' r'^2 \int\limits_{\frac{r'}{2d}} ^{1} dy' T_{\gamma} c_{\gamma} ({\bf r}_1+{\bf r}') \nonumber\\
\end{align}
where $y'=\cos \theta'$. Finally, by Taylor expanding the concentrations around ${\bf r}_1$ up to first order in displacement vectors, we obtain:

\begin{align}
\nabla_{{\bf r}_1}\Phi_{\alpha}^{(2)}=&\sum_{\beta,\gamma} \bigg[  T_{\gamma}  \int \widehat{{\bf r}} \hspace{2pt} \delta 
(r-d)   \\ &\nonumber
 \int\limits_{V_{\cap}}\big[ c_{\beta} ({\bf r}_1)c_{\gamma} ({\bf r}_1)  +  c_{\gamma} ({\bf r}_1) \hspace{2pt}{\bf 
 r}\cdot\nabla_{{\bf r}_1} c_{\beta} ({\bf r}_1) \\ &\nonumber \quad+ c_{\beta} ({\bf r}_1) \hspace{2pt}{\bf r'} \cdot 
 \nabla_{{\bf r}_1} c_{\gamma} ({\bf r}_1) \big]  d {\bf r}' d {\bf r}\bigg].
\end{align}
Clearly, the first term, i.e., $c_{\beta} ({\bf r}_1)c_{\gamma} ({\bf r}_1) $ vanishes upon integration. The second 
term can be integrated by writing $\nabla_{{\bf r}_1} c_{\gamma} ({\bf r}_1)\equiv \widehat{{\bf z}} |\nabla_{{\bf r}
_1} c_{\gamma} ({\bf r}_1)| $. A similar implementation on the third term (using 
product rules) gives:

\begin{align}
\nabla_{{\bf r}_1}\Phi_{\alpha}^{(2)}=C \sum_{\beta,\gamma} T_{\gamma}\big [&2 c_{\gamma} ({\bf r}_1) \nabla_{{\bf r}_1} 
c_{\beta} ({\bf r}_1) \nonumber\\& + c_{\beta} ({\bf r}_1) \nabla_{{\bf r}_1}  c_{\gamma} ({\bf r}_1) \big].
\end{align}
For equilibrium systems with $\ma{T}=\mb{T}$, this result gives back the third-virial coefficients between 
hard-spheres. In the general case, for mixtures of particles with two different 
temperatures, we obtain Eq.\eqref{pressure3} of the main text
and $\nabla \Phi_{\mathcal{B}}^{(2)}=\nabla \Phi_{\mathcal{A}}^{(2)}$. Hence, it appears that this field is 
not integrable to obtain $\Phi_{\mathcal{B}}^{(2)}$. On the other hand, we observe that $\ma{c}\nabla 
\ma{\mu}+\mb{c}\nabla \mb{\mu}=\nabla p$ consistent with the result obtained from virial equation \eqref{virpress}. 

This method could as well be implemented for mixtures of hard-spheres with different diameters $\ma{d}\neq 
\mb{d}$ as well as different temperatures $\ma{T}\neq \mb{T}$. This would require to solve the general 
form of  Eq.\eqref{phieq} with varying contact distances,

\begin{align}\label{geneq}
\nabla_{{\bf r}_1}\Phi_{\alpha}^{(2)}&=\sum_{\beta,\gamma} \int \widehat{{\bf r}} \hspace{2pt}\delta (r-d_{12}) c_{\beta} ({\bf r}_1 + {\bf r}) \\& \nonumber  \int T_{\gamma} c_{\gamma} ({\bf r}_1+{\bf r}') \Theta (d_{13}-r') \Theta (d_{23}-|{\bf r}- {\bf r}'|)  d {\bf r}' d {\bf r}.
\end{align}
\subsection{Phase separation and coexistence conditions}
Following the procedure described in Section 3, we determine the phase diagrams at third-order 
in concentration. A first remark is that we find an instability when $\ma{T}/\mb{T}\gtrapprox \rev{6.171}$ for equal-sized 
spherical particles. Hence, the third order terms delay the onset of phase separation compared to the 
second order expansion which leads to phase separation when $\ma{T}/\mb{T}> 4 $ \cite{Note2}.
The phase coexistence conditions can be calculated by imposing that the particle fluxes to 
zero as well as the momentum flux. This last condition is obtained from  $\ma{c}\nabla \ma{\mu}+\mb{c}
\nabla \mb{\mu}=\nabla p=0$, which gives mechanical equilibrium,  and therefore imposes that 
pressure is constant. The other 
conditions require that $\nabla \ma{\mu}=0$ and $\nabla \mb{\mu}=0$. However, 
the chemical potential gradients are 
not integrable. Nevertheless, since $\nabla \ma{\Phi}^{(2)}=\nabla \mb{\Phi}^{(2)}$, we can use the 
condition that $\nabla (\ma{\mu}-\mb{\mu})=0$ which would lead to two gradient free equations. Therefore, to
summarize, we find two conditions for the the $a$ and $b$ phases to coexist 
\begin{eqnarray}
\ma{\mu}^{(1)}\left(\ma{c}^{a},\mb{c}^{a}\right)-\mb{\mu}^{(1)}\left(\ma{c}^{a},\mb{c}^{a}\right)&=&\ma{\mu}^{(1)}(\ma{c}^{b},\mb{c}^{b})-\mb{\mu}^{(1)}(\ma{c}^{b},\mb{c}^{b})\nonumber\\ \\
p\left(\ma{c}^{a},\mb{c}^{a}\right)&=&p(\ma{c}^{b},\mb{c}^{b}).
\end{eqnarray}
where we defined with $\mu_{\alpha}^{(1)}=\mu_{\alpha}^{\text{id}}+\Phi_{\alpha}^{(1)}$, $\alpha=\mathcal{A,B}$.

The third condition required to construct the phase diagrams is not directly accessible in a general 
form due to the  lack of well-defined chemical potentials. Yet, it is possible to derive an approximate 
condition in the vicinity of the 
critical point by linearizing the time evolution equations in concentrations.

\bibliography{mainv1}

\end{document}